\documentclass[final,5p,times,twocolumn]{elsarticle}

\usepackage{amssymb}
\usepackage{amsfonts}
\usepackage{amssymb}
\usepackage{amsmath}  
\usepackage{amsthm}
\usepackage{mathtools}
\usepackage{enumitem} 
\usepackage{geometry}
\usepackage{graphicx}
\usepackage{algorithm}
\usepackage{algcompatible}
\usepackage[colorlinks=true, allcolors=blue]{hyperref}
\usepackage{hyperref}
\usepackage[utf8]{inputenc} 
\usepackage[english]{babel} 
\usepackage{url}
\usepackage{wrapfig}
\usepackage{subfiles}
\usepackage{xfrac}
\usepackage{multirow}
\usepackage{cleveref}
\usepackage{balance}
\usepackage{booktabs}
\usepackage{tikz}
\usetikzlibrary{shapes.geometric, arrows}
\usepackage{tikz-3dplot}
\usepackage[squaren]{SIunits}
\usepackage{caption}
\usepackage[normalem]{ulem}
\usepackage{subcaption}
\usepackage[figuresleft]{rotating}
\usepackage{lipsum}
\usepackage{todonotes}
\usepackage{float}
\usepackage{pgfplots}
\pgfplotsset{compat=newest}

\usepackage{pgffor} 

\newtheorem{proposition}{Proposition}
\newtheorem{assumption}{Assumption}
\newcommand{\bc}{\begin{center}}
\newcommand{\ec}{\end{center}}
\newcommand{\be}{\begin{equation}}
\newcommand{\ee}{\end{equation}}
\newcommand{\bea}{\begin{eqnarray}}
\newcommand{\eea}{\end{eqnarray}}
\newcommand{\beq}{\begin{eqnarray*}}
\newcommand{\eeq}{\end{eqnarray*}}
\newcommand{\bv}{\left( \begin{array}{c} }
\newcommand{\ev}{\end{array} \right) }

\newcommand{\E}{\mathbb{E}}

\newcommand{\Var}{\mbox{Var}}

\newcommand{\lt}{\ensuremath <}

\makeatletter
\def\ps@pprintTitle{%
  \let\@oddhead\@empty
  \let\@evenhead\@empty
  \def\@oddfoot{\reset@font\hfil\thepage\hfil}
  \let\@evenfoot\@oddfoot
}

\usepackage{natbib}

\begin{document}

\begin{frontmatter}

\title{The bias of IID resampled backtests for rolling-window mean-variance portfolios}

\author[first]{Andrew Paskaramoorthy}
\affiliation[first]{organization={Department of Statistical Sciences},
            addressline={University of Cape Town}, 
            city={Cape Town},
            postcode={7701}, 
            state={Western Cape},
            country={South Africa}}
\author[second]{Terence van Zyl}
\affiliation[second]{organization={Institute for Intelligent Systems},
            addressline={University of Johannesburg}, 
            city={Johannesburg},
            postcode={}, 
            state={Gauteng},
            country={South Africa}}
\author[first]{Tim Gebbie }

\begin{abstract}
Backtests on historical data are the basis for practical evaluations of portfolio selection rules, but their reliability is often limited by reliance on a single sample path. This can lead to high estimation variance. Resampling techniques offer a potential solution by increasing the effective sample size, but can disrupt the temporal ordering inherent in financial data and introduce significant bias. This paper investigates the critical questions: First, \emph{How large is this bias for Sharpe Ratio estimates?}, and then, second: \emph{What are its primary drivers?}. We focus on the canonical rolling-window mean-variance portfolio rule. Our contributions are identifying the bias mechanism, and providing a practical heuristic for gauging bias severity. We show that the bias arises from the disruption of train-test dependence linked to the return auto-covariance structure and derive bounds for the bias which show a strong dependence on the observable first-lag autocorrelation. Using simulations to confirm these findings, it is revealed that the resulting Sharpe Ratio bias is often a fraction of a typical backtest's estimation noise, benefiting from partial offsetting of component biases. Empirical analysis further illustrates that differences between IID-resampled and standard backtests align qualitatively with these drivers. Surprisingly, our results suggest that while IID resampling can disrupt temporal dependence, its resulting bias can often be tolerable. However, we highlight the need for structure-preserving resampling methods.
\end{abstract}


\end{frontmatter}




\section{Introduction}


Given some portfolio rules, which portfolio rule performs the best? It would be ideal if we could merely compare the theoretical performance across these rules; however, this is typically intractable. This leaves us to rely on performance estimates. For this reason, ``Backtests'', {\it i.e.} out-of-sample (OOS) performance estimates on historical data, are ubiquitous in finance and have been used to make important claims in the portfolio selection literature. 


For example, \citet{BM2009} compares several portfolio rules' empirical performance on multiple datasets and concludes that an optimised portfolio typically fails to outperform a naively diversified ({\it  i.e.} equally weighted) portfolio. However, a closer inspection of these results shows that performance differentials often appear large but are not statistically significant. The conclusions of \citet{BM2009} may, and seem, to remain valid. However, this disconnect appears to highlight the inadequacy of our current statistical tests to portray an otherwise clear empirical story. 

At first glance, the main problem appears to be a lack of data -- at least if one assumes that the world is not changing very quickly and that past data is reasonably representative of future outcomes. Conventional backtests retain the temporal ordering of data when splitting data into training (in-sample) and validation (out-of-sample) data. If a sample path of length $T$ is split into a training set of $n$ observations, $T-n$ out-of-sample observations are used for performance estimation. When $n$ is large relative to $T$, as for large complex models, the lack of data and resulting performance estimation inaccuracy is acute. Surprisingly, even for small models where $n \ll T$, the lack of data remains a problem, particularly for non-Gaussian temporally dependent data. A back-of-the-envelope calculation using the asymptotic distribution of the Sharpe Ratio shows that under fairly typical return moments \cite{eling2010performance}, detecting an annual Sharpe Ratio difference of $0.5$ between two competing strategies, with $80\%$ success, requires approximately $400$ months of data. Tests either have very low power or require impractically large sample sizes to detect out-performance -- this is even before one considers path dependencies and the dynamic changes in the underlying reality of the data-generating process itself.

If naive re-ordering of the data were feasible, the lack of data could be resolved through resampling as if the data were IID (Independent and Identically Distributed). At an extreme, if we could assume IID observations, we could form $\binom{T}{n} \times \left(T-n\right)$ train-validation pairs from the sample path, producing a substantial reduction in the performance statistic's standard error, with a notable increase in the power of any accompanying test\footnote{Since data are reused for training and validation, the out-of-sample observations are not independent. Although $\binom{T}{n}$ may be very large, the standard error of CV estimators will often not vanish}. However, IID resampling is seen as problematic primarily because it would disrupt the dynamic and sometimes transient temporal structure inherent in financial data. 

Asset returns are time-varying, displaying local and global structure. Local dependence arises from slow variation in parameters typical of normal economic activity. Global structure arises from sudden regime shifts often precipitated by global market events, creating dislocations in return distributions over time. However, structural changes can be confused with long-memory, and these can mix both local and global structures dynamically through time. Locally, shuffling data would disrupt dependence within the training set itself and between the training and validation sets. Globally, IID resampling may select training and validation data from vastly different distributions reflecting radically different market conditions. Lastly, the failure to mimic real-world implementation, particularly using future data for training, is unsettling for many given the reflexive nature of real financial markets. 

Yet, the impact of re-ordering data for performance estimation is not broadly established. Raboniwicz and Rosset \cite{rabinowicz2022cross} prove that if a CV method splits data to preserve the joint distribution of training and test data under real-world implementation, the CV method will be unbiased. However, the converse is not necessarily true. For example, Bergmeir \cite{bergmeir2018note} proves that k-fold CV is a consistent estimator under Auto-Regressive data (for large models). Nonetheless, re-ordering dependent data often results in biased performance estimation, motivating block-resampling methods that preserve local dependence \cite{de2018advances} -- but potentially compromise long-memory affects. Furthermore, whether there is a performance estimation benefit to re-ordering averaged (smoothed) or down-sampled (decimated) data, rather than to average, and then resample; or resample and then average (or sub-sample); the empirical impact of these types of choices remains unclear.  

IID resampling has not been extensively studied, if at all, for small sample models using rolling window estimates. We aim to quantify the bias and variance of an IID-resampled backtest performance estimator and establish conditions where IID resampling could be useful for estimating the unconditional performance of small-model mean-variance portfolio rules. Our study is motivated by a simple observation as part of a preliminary exploratory data analysis. Namely, the dynamics of the rolling mean estimates are governed by estimation error and time variation of the underlying risk premium. As noise becomes increasingly large, estimation errors dominate the estimates, irrespective of temporal structure in the risk premium. Motivated by this observation, we hypothesize that:
\textit{The bias of IID resampled performance estimators depends on the variance of the time-varying risk premium relative to the total variance, i.e. the Signal-to-Noise Ratio (SNR), and the bias from IID resampling is negligible under low SNR market conditions.} 



\subsection{Contributions}

Our main contribution is to quantify the bias of IID resampled Sharpe Ratio estimators for rolling-window mean-variance portfolio rules, eliciting conditions where IID resampling may be feasible. Along with the naively diversified portfolio, the rolling-window mean-variance rule is a standard benchmark in the portfolio selection literature.

Analytically, we derive bounds for the bias of the IID resampled Sharpe Ratio, demonstrating its dependence on the first-lag autocorrelation of the returns and the Sharpe Ratio of the risky asset. The first-lag autocorrelation of returns depends on the SNR and the persistence of the risk premium of the underlying asset's return processes. This supports our hypothesis that the bias will also be low when the SNR is low. For realistic return autocorrelation, the true bias is substantially smaller than the bound, suggesting that the IID resampling can be feasible even when the bias is large.

The remainder of this paper is organised as follows. In Section \ref{sec:formulation}, we define the primary object of our study: the IID resampled backtest estimator's bias. In Section \ref{sec:biasbounds}, we present our main analytical results, expressing bounds for the bias of the mean, variance, and Sharpe Ratio of the IID resampled estimator as a function of parameters of the data-generating process. In Section \ref{sec:simulation}, we present the results of our simulation experiments. In Section \ref{sec:historical}, we present the results using real-world data. Finally, in Section \ref{sec:conclusion}, we summarise our findings and discuss future avenues of research.


\section{Problem Formulation}\label{sec:formulation}

We start by defining a portfolio rule as a statistical decision rule; a function that takes in data $S_t$ at each period $t$ and outputs a portfolio: 
$$
\delta(S_t) = w_t.
$$
In this paper, we focus specifically on the canonical rolling-window mean-variance (MV) portfolio rule, where $w_t$ depends on the sample mean and variance calculated over the historical data: $S_t = \{R_{t-n+1},\ldots, R_t \}$.

The portfolio's performance is measured by its realised return over the subsequent period:
$$
	R_{p,t+1} = w_t'R_{t+1}
$$
where $R_{t+1}$ is a vector of out-of-sample returns for each asset in the portfolio. 

\subsection{Portfolio Rule Performance}
The portfolio rule's performance is defined by the moments of its realised return distribution. We focus on the unconditional performance of applying the portfolio rule repeatedly over time, which we measure by the unconditional Sharpe Ratio, defined by
$$
\Theta_p  = \frac{\mu_{p}}{\sigma^2_{p}} = \frac{\E[R_{p,t+1}]}{\Var(R_{p,t+1})},
$$
where the expectations are taken over all random quantities. For convenience, we have used that $\mu_{p} = \E[R_{p, t+1}]$ and $\sigma^2_{p} = \Var(R_{p, t+1})$. We assume that the portfolio rule's out-of-sample returns are stationary. 

\textit{Remark:} To ensure these unconditional moments are well-defined characteristics of the portfolio rule's long-run performance, we assume that the realised return process is stationary and ergodic\footnote{Here $\E[R_t | {\cal F}_{t-\tau}]= h(R_{t-\tau})$ for some time $t$, lag $\tau$, filtration $\cal F$ and smooth function $h$, and that $\lim_{\tau \to \infty} \E[R_t | {\cal F}_{t-\tau}]= \E[R_{t}]$. }. While stationarity is a strong assumption for financial returns that exhibit time-variation, long-memory, and potential regime shifts, it is a standard assumption for analytical work {\it e.g.} see \cite{lo2002statistics, opdyke2007comparing}). We adopt this assumption in our framework, acknowledging the potential impact of non-stationarities like regime changes on the interpretation of standard backtests and the derived bias, which we discuss later, particularly in the context of the empirical work.

\subsection{Standard Backtests}
A standard backtest involves applying the portfolio rule sequentially over a historical sample path. In particular, for a sample path of length $T$ and a rolling window size of $n$, we apply the portfolio rule at each period $t = n, \dots, T-1$, and measure its realised return over the subsequent period $t+1$, yielding the sequence $\lbrace R_{p,n+1}, \dots, R_T \rbrace$. The portfolio's Sharpe Ratio is estimated from the $N=T-n$ realised returns using the standard sample moment estimators: 
$$
	\widehat{\Theta}_p:=\widehat{\Theta}_p\left(R_{p,T-n+1}, \dots, R_{p,T}\right) = \frac{\widehat{\mu}_p}{\widehat{\sigma}_p}.
$$
Here, the portfolio rule's estimated mean return and standard deviation given $N=T-n$ are:
$$
\widehat{\mu}_p = \frac{1}{N}\sum_{i=n+1}^{T}R_{p,i} \quad \text{and} \quad \widehat{\sigma}_p = \sqrt{\frac{1}{N-1}\sum_{i=n+1}^{T}\left(R_{p,i} - \widehat{\mu}_p\right)^2}.
$$ 

Assuming stationarity, the backtested Sharpe Ratio is an unbiased and consistent estimator of the true Sharpe Ratio, and the Central Limit Theorem then gives the asymptotic distribution \cite{lo2002statistics}:
$$
	\frac{\widehat{\Theta}_p - \Theta}{\ \text{SE}\left(\widehat{\Theta}_p\right)} \xrightarrow{d} N\left(0,  1 \right),
$$
where the Standard Error (SE) is inversely proportional to the square root of the sample size $\text{SE}\left(\widehat{\Theta}_p\right) \propto N^{-1/2}$. 

\subsection{IID Resampled Backtests}
An IID Resampled backtest involves applying IID resampling without repetition ({\it i.e.}, shuffling) to the historical sample path, and then calculating a standard backtest estimate on the resampled path. 

Formally, we form a resampled sample path of length $T$ (the same length as the original sample path):
$$
\left({R^{*}_{1}, R^{*}_2, \dots, R^{*}_T }\right)
$$
where $R^{*}_{t}$ denotes a resampled observation. As before, we apply the portfolio rule at each period $t=n, \dots, T-1$ and then calculate its realised return at time $t+1$ using the resampled sample path:
$$
\begin{aligned}
\delta \left(S^{*}_t\right) &= w^{*}_t \\
R^{*}_{p,t+1} &= {w^{*}}_t' R_{t+1}^{*}.
\end{aligned}
$$
Lastly, we calculate the resampled backtest Sharpe Ratio from the sequence of realised returns:
$$
	\widehat{\Theta}_p^{*}:=\widehat{\Theta}_p\left(R^{*}_{p,n+1}, \dots, R^{*}_{p,T}\right)
$$
IID resampling makes the asset returns IID. Therefore, the CLT also gives the asymptotic distribution of the backtested Sharpe Ratio after resampling, except with parameters based on IID data. We have:
$$
	\E\left[\widehat{\Theta}_p^{*}\right] = \Theta_p^{*} \quad \text{and} \quad \widehat{\Theta}_p^{*} \rightarrow \Theta^{*}_p.
$$
where $\Theta^{*}_p$ denotes the portfolio rule's Sharpe Ratio measured on IID asset returns, where the asset returns have the same unconditional moments as the original DGP. 

Although the asset returns are IID, some autocorrelation will remain in the portfolio rule's out-of-sample returns due to overlapping training sets from the rolling windows. Nonetheless, the standard error of the IID-resampled Sharpe Ratio will likely be lower than the original Sharpe Ratio.

\subsection{The Bias of IID Resampled Backtest Sharpe Ratios}

The bias of the resampled Sharpe Ratio estimator is given by:
$$
\text{bias}\left(\widehat{\Theta}_p^{*}\right) = \Theta_p^{*} - \Theta_p.
$$
We can conveniently express the bias of the resampled Sharpe Ratio in terms of the bias of the resampled mean and variance using a first-order Taylor approximation. 

Expanding $\Theta_p = \hat{\mu}_{p} \left(\hat{\sigma}_p^{2}\right)^{-1/2}$ around the portfolio rule's IID resampled mean and variance $(\mu^{*}_p, \sigma^{*2}_{p})$ gives:

\begin{equation}\label{eq:taylor}
\Theta_p \approx \tfrac{\mu^{*}_p}{\sigma^{*}_p} + \tfrac{1}{\sigma^{*}_p}\left({\mu}_p - \mu^{*}_p\right) - \tfrac{\mu^{*}_p}{2\sigma^{*3}_p}\left({\sigma}_p^{2}-\sigma^{*2}_p\right).
\end{equation}

By subtracting \ref{eq:taylor} from $\Theta^{*}_p$ and dividing by $\Theta_p$, we can express the relative bias of the resampled Sharpe ratio in terms of the relative bias of the resampled mean and variance.


\begin{equation}\label{eq:bias_sharpe}
\tfrac{1}{\Theta_p^{*}}\text{bias}\left(\widehat{\Theta}_p^{*}\right) \approx \tfrac{1}{\mu^{*}_{p}}\text{bias}\left(\widehat{\mu}^{*}_p\right) -\tfrac{1}{2 \sigma^{*2}_{p}}\text{bias}\left(\widehat{\sigma}_p^{*2}\right)
\end{equation}

The absolute bias can be expressed by multiplying both sides of Equation \ref{eq:bias_sharpe} by $\Theta^{*}_p$.


\section{The Size of the Bias}\label{sec:biasbounds}

\subsection{Assumptions}\label{ssec:assumptions}

We consider the problem of allocating between a single $M=1$ risky asset and a risk-free asset using mean-variance optimisation. For analytical tractability, we make the following assumptions.

\begin{assumption}\label{ass1}
    The covariance of the risky assets $\Sigma$ is known and is constant.
\end{assumption}

\begin{assumption}\label{ass2}
Forecasts for the expected return $\mu_{t+1}$ are formed using the simple rolling average of past returns with a window size $n$:
$$
\hat{\mu}_{t+1} = \frac{1}{n}\sum_{i=t-n+1}^{t}R_{i}. 
$$
\end{assumption}

Therefore, the mean-variance optimised portfolio output by our portfolio rule at each period $t$ is given by:
$$
w_{t} = \frac{1}{\gamma} \Sigma^{-1}\hat{\mu}_{t}
$$
where $\gamma$ is the investor's risk aversion.

\begin{assumption}\label{ass3}
The risky asset returns $R_t$ are generated by a stationary process with constant variance, where the time-varying means $\mu_t$ follow a AR(1) process:

$$
\begin{aligned}
	R_t &= \mu_t + \epsilon_t \\
	\mu_t &= \mu_0+\phi\mu_{t-1} + \eta_t.	
\end{aligned}
$$

where $R_{t},\mu_{t}$, $\epsilon_{t}$ and $\eta_{t}$ are $m \times 1$ vectors, with $\epsilon_{t} \sim N(0,\sigma^2_{\epsilon})$ and $\eta_{t} \sim N(0,\sigma^2_{\eta})$ independently, with auto-correlation $\phi$. 
\end{assumption}
Temporal dependency is expressed by the auto-correlation parameter with the standard restriction to ensure stationarity $|\phi|< 1$. Furthermore, we have that the total variance can be decomposed into the variance of the risk premium and the variance of the noise: $\sigma^2_{R}=\sigma^2_{\mu}+\sigma^2_{\epsilon}$. The variance of the risk premium depends on the autocorrelation and the noise in the risk premium process: 
\begin{equation}\label{eq:sigma_mu}
\sigma_{\mu}^{2} = \frac{\sigma_{\eta}^{2}}{1-\phi^{2}}.
\end{equation}



An important consequence of Assumption \ref{ass3} is that the first lag autocorrelation of the asset returns is given by:
\begin{equation}\label{eq:psi}
    \psi := \operatorname{Corr}\left(R_t,R_{t+1}\right) = \phi\tfrac{\sigma_{\mu}^2}{\sigma_R^2}
\end{equation}
The first lag autocorrelation is a key variable to characterise the bias of the resampled backtests. 

\textit{Remark 1:} The asset return process corresponds to a simplified conditional factor model:
\begin{align}
R_t &= \beta f_{t}+\epsilon_t, \nonumber \\
f_t &= \gamma_t + v_t. \nonumber
\end{align}
Here $\beta_t$ are static factor exposures, $f_t$ are the factors, $\gamma_t$ is a time-varying factor premium, and $v_t$ are IID factor innovations. Assuming that the factor risk premium follows an AR(1) process $\gamma_t = \gamma_0 + \phi\gamma_{t-1} + \eta_t$, the asset pricing model becomes
$$
\begin{aligned}
R_t &= \tilde{\mu}_t+\tilde{\epsilon_t} \\
\tilde{\mu}_t &= \tilde{\mu}_0+ \tilde{\mu}_{t-1} + \tilde{\eta}_t
\end{aligned}
$$
where $\tilde{\mu}_t = \beta\gamma_t$, $\tilde{\eta}_t=\beta\eta_t$ and $\tilde{\epsilon}_t = \beta v_t+\epsilon_t$. Thus, the noise term includes both idiosyncratic and factor risk. 

\textit{Remark 2:} Our assumed DGP specifies a smoothly time-varying risk premium but omits conditional heteroskedasticity -- a well-established feature of empirical returns. We omit conditional heteroskedasticity in our model for analytical tractability, but study its effects in our simulation experiments. Notably, our results suggest that conditional variance does not materially affect the bias in the Sharpe Ratio (see Section \ref{ssec:results}).

\textit{Remark 3:} Our model specifies a smoothly time-varying risk-premium. While time variation in risk premia is well established in the literature, the dynamics of risk premia are less studied and appear to depend on the estimation methodology. Results from \cite{adrian2015regression} and \cite{gagliardini2016time} suggest that factor-risk premia are smoothly varying and highly persistent, with greater variation during market-crisis periods. Yet, factor returns exhibit first-lag auto-correlations of $0.1$ \cite{gupta2019factor} on average, indicating that alternative sources of variation highly dampen the persistence exhibited by the risk premium as it propagates to the returns. In our model, we can simultaneously capture the empirical phenomena of highly persistent risk premia with low persistence in returns by controlling the noise covariance. On the other hand, alternative evidence suggests that temporal variation in factor risk-premia is better explained by discrete jumps representing regime changes \cite{pastor2001equity,smith2022have}. We discuss the implications of regime changes on our analytical results in Section \ref{sec:discussion}.

\subsection{The Bias of the Resampled Mean}

The mean return of a portfolio rule can be decomposed into two terms:
\begin{align}\label{eq:mean_ret}
    \mu_p &= \E[w_tR_{t+1}]= \E[w_t]E[R_{t+1}] + \text{cov}\left(w_t,R_{t+1}\right).
\end{align}
The second term, represented by the covariance between the portfolio weights and future returns, reflects the temporal relationship between the training and test data. Since IID resampling eliminates all temporal dependency in the underlying assets' returns, the covariance term in Equation \ref{eq:mean_ret} becomes zero for an IID-resampled backtest. 

IID resampling does not alter the expected return of the underlying asset. And, assuming that the covariance matrix is known and constant (Assumption $\ref{ass1}$), IID resampling does not alter the expected portfolio. Hence, the bias of the mean return for an IID-resampled backtest of an MV portfolio rule is given by the covariance term:
\begin{equation}
\begin{aligned} \label{eq:biasm}
\text{bias}\left(\hat{\mu}^{*}_p\right) &= -\text{cov}\left(w_t,R_{t+1})\right) = -\tfrac{1}{\gamma\sigma_{R}^{2}}\text{cov}(\hat{\mu}_{t},R_{t+1}).     \\
\end{aligned}    
\end{equation}

From Equation \ref{eq:biasm}, it is straightforward to prove the following proposition (see Appendix \ref{appsec:mean_bound}):
\begin{proposition}\label{prop:bias-mean}
Under assumptions \ref{ass1} - \ref{ass3}, and assuming that $0\leq \phi < 1$, the bias of the IID-resampled backtested mean of a mean-variance portfolio is given by:
\begin{equation}
-\tfrac{1}{\gamma}\psi\leq -\tfrac{1}{\gamma}\tfrac{\sigma^2_{\mu}}{\sigma_{R}^{2}}A_{n,\phi} = \text{bias}\left(\hat{\mu}_{p}^{*}\right) \leq 0
\end{equation}
where $A_{n,\phi} = \frac{1}{n}\sum_{i=1}^n \phi^{t+1-i}$. 
\end{proposition}

\textit{Remark:} Recall from Equation \ref{eq:sigma_mu}, that $\sigma^2_{\mu}$ depends on $\phi$. If $\phi=1$, the risk-premium follows a random walk and its variance $\sigma^2_{\mu}$ is not defined. 

Proposition \ref{prop:bias-mean} states that the bias of the IID-resampled backtest mean is negative (underestimates the true mean). The magnitude of the bias is given by the product of the reciprocal of the risk aversion parameter, the ratio of the risk premium's variance to the return variance, and the sum $A_{n,\phi}$. The risk aversion parameter controls the size of the portfolio. The ratio of the risk premium's variance to the return variance reflects the ``signal" content of the returns. The sum $A_{n,\phi}$ captures the dependence between the training and test data for the rolling mean forecast. 

Note that for the rolling mean forecast, the sum $A_{n,\phi}$ depends on the persistence of the risk premium $\phi$ and the window size $n$. For a fixed $\phi$, the sum $A_{n,\phi}$ is monotonically decreasing with $n$ with upper and lower limits: 
$$
\begin{aligned}
n=1: &A_{n,\phi} = \phi, \quad \text{and}    \\
\lim_{n \to \infty} &A_{n,\phi} = 0
\end{aligned}
$$
Under our assumed DGP, autocorrelation is greatest in the short term. As the window size $n$ increases, the autocorrelation is averaged over longer periods, reducing the impact of short-term autocorrelation on the portfolio rule's performance, hence decreasing the portfolio rule's mean-return. 

The parameters $\sigma_\mu$ and $\phi$ are difficult to estimate, and the ranges for these parameters are not commonly reported. However, using Equation \ref{eq:psi}, we can express the bound of the bias in terms of an observable quantity, the lag-1 autocorrelation of the asset returns $\psi$: 
\begin{equation}
    \psi :=\text{corr}\left(R_{t}, R_{t+1}\right) =  \tfrac{\sigma^2_{\mu}}{\sigma_{R}^{2}}\phi
\end{equation}
This is useful as it states an empirically observable condition for when IID resampling could be dangerous: a ``low" $\frac{1}{\gamma} \psi$ indicates that bias will also be low. The notion of ``low" is a subjective judgement of the backtester. In Section \ref{ssec:results}, we compare magnitudes relative to standard errors for context. 

\subsection{The Bias of the Resampled Variance}

Once again, the variance of the portfolio rule can be broken down into three terms (derivation in Appendix \ref{appsec:var_bound}):
$$
\begin{aligned}\label{eq:var}
	\sigma_p^2 = \E \left[ w_t \right]^{2}\text{var}\left( R_{t+1} \right) + \E \left[ R_{t+1}^{2} \right]\text{var}\left(w_{t}  \right) + \operatorname{cov}\left(w_{t}^{2},R_{t+1}^{2}\right).
\end{aligned}
$$
The first term reflects the variation arising from the asset returns, scaled by the expected portfolio, and is invariant under IID resampling. The second term reflects the variation from estimating the portfolio, a function of the dependency within the training data. The third term captures effects related to a higher-order temporal dependency between the portfolio weights and the future returns. 

IID resampling breaks the return dependence, affecting the second and third terms in Equation \ref{eq:var}. Accordingly, the bias of the IID resampled backtest variance is given by (derivation in Appendix \ref{appsec:var_bound}):
\begin{equation}\label{eq:biass}
\text{bias}\left(\hat{\sigma}_p^{*2}\right) = \E \left[ R_{t+1}^{2} \right]\left({\text{var}(w^{*}) - \text{var}(w)}\right) - \tfrac{1}{2}\operatorname{cov}\left(w^{2},R_{t+1}^{2}\right).
\end{equation}
The first term in Equation \ref{eq:biass}, comprised of the difference of variances, arises from removing the dependence within the training set during IID resampling. The second term arises from removing the dependence between the training and test sets. 

The covariance term captures a higher-order dependence between the portfolio weights and the future returns. However, suppose the weights and the future returns have a joint Gaussian distribution. In that case, the second-order dependence can be described in terms of first-order dependence, implying that the bias of the variance depends on the bias of the mean. This is important since the bias of the Sharpe Ratio offsets the bias of the mean and the variance.   

From Equation \ref{eq:biass}, we prove Proposition \ref{prop:bias-var}.
\begin{proposition}\label{prop:bias-var}
Under assumptions \ref{ass1} -- \ref{ass3}, and assuming $0 \leq \phi < 1$, the bias of the IID-resampled backtested variance of a mean-variance portfolio is given by:
\begin{equation}
    \text{bias}\left(\hat{\sigma}_p^{*2}\right) = -\tfrac{1}{\gamma^{2}} \left[{ \left(\theta^{2}+1  \right) \tfrac{\sigma_{\mu}^{2}}{\sigma_{R}^{2}} B_{n,\phi} + \tfrac{\sigma_{\mu}^{2}}{\sigma_{R}^{2}}A_{n,\phi}\left( \tfrac{\sigma_{\mu}^{2}}{\sigma_{R}^{2}}A_{n,\phi} + 2\theta^{2} \right) }\right]
\end{equation}
where $B_{n,\phi} = \frac{1}{n^2} \sum_{i}\sum_{j\neq i}\phi^{|i-j|}$, and $A_{n,\phi}$ is defined as before. With $\theta$, the risky asset's Sharpe Ratio, the bias can be bounded:
\begin{equation}
-\tfrac{C}{\gamma^{2}}\psi < \text{bias}\left(\hat{\sigma}_p^{*2}\right) \leq 0,
\end{equation}
where $C$ is the polynomial $C = 3\theta^{2} + \psi + 1$.
\end{proposition}
The bias has two components, where the term involving $B_{n,\phi}$ reflects the bias arising from breaking dependence within the training set, and the term involving $A_{n,\phi}$ reflects the portion of the bias from breaking the dependence between the training and test sets. The sums $A_{n,\phi}$ and $B_{n,\phi}$ are specific to the rolling mean forecast, and would be defined differently for a different forecast rule. Similarly, $B_{n,\phi}$ also depends on risk premium persistence $\phi$ and window size $n$ with similar limiting behaviour as $n$ increases. 

The bias bound for the resampled variance estimator resembles the bound for the resampled mean. In particular, the bias for the variance is negative and its bound depends on the reciprocal of the squared risk aversion $\gamma$, and the first-order autocorrelation of the returns $\Psi$. However, in contrast to the bound for the mean returns, the bound for the variance depends on an additional scaling term $C$, a polynomial that is a function of the assets' Sharpe Ratio and autocorrelation.  Nonetheless, although the bound contains the additional term $C$, numerical experiments (not shown) indicate that the lag-1 autocorrelation dominates the bound.

\subsection{The Bias of the Sharpe Ratio}\label{ssec:biasbounds}

Equation \ref{eq:bias_sharpe} shows that the bias of the Sharpe Ratio offsets the relative bias of the mean with the relative bias of the variance. For the rolling window MV portfolios we study here, we can further simplify this relationship by substituting the moment approximations $\mu_p^{*} \approx \left(1/\gamma\right)\theta^2$ and $\sigma_p^{*2} \approx \left(1/\gamma^2\right)\theta^2$ (Proofs in Appendix \ref{appsec:mean_bound} and \ref{appsec:var_bound}) into Equation \ref{eq:bias_sharpe}:
\begin{equation}\label{eq:mv_sharpebias}
    \text{bias}\left(\widehat{\Theta}_p^{*}\right)  \approx \tfrac{\gamma}{\theta} \left(\text{bias}\left(\widehat{\mu}^{*}_p\right) -\tfrac{\gamma}{2}\text{bias}\left(\widehat{\sigma}_p^{*2}\right) \right),
\end{equation}
where $\theta$ is the Sharpe Ratio of the underlying asset. Equation \ref{eq:mv_sharpebias} expresses the bias in the Sharpe Ratio in terms of the difference between the bias in the mean and the variance, and the asset's Sharpe Ratio.

Crucially, Equation \ref{eq:mv_sharpebias} demonstrates the offsetting effect: the contribution of the bias from the mean is offset by the bias in the variance, if these component biases move in tandem with one another. Conversely, if these components move in opposition to one another, the bias of the SR can increase. Thus, the bias of the SR depends heavily on the relationship between these component biases. Under Assumptions \ref{ass1} -- \ref{ass3}, the component biases are positively nonlinearly related for the MV portfolio rule for finite window size $n$. Defining $R_n = B_{n,\phi}/A_{n,\phi}$, we show that 
$$
    R_n = 2\left(\frac{1}{1-\phi^n} - \frac{1}{n(1-\phi)}\right).
$$

However, as we prove in Appendix \ref{appsec:assymptotic_sums}, the component biases are approximately linearly related, especially as window size $n$ increases: 
$$
    \lim_{n \to \infty} R_{n} = 2.
$$
This asymptotic linearity facilitates the offsetting mechanism, often resulting in an SR bias that is smaller in relative magnitude than either component bias.

Direct substitution of the full expressions for the component biases (Propositions \ref{prop:bias-mean}) and \ref{prop:bias-var}) into Equation \ref{eq:mv_sharpebias} and collecting terms  yields a complicated analytical expression for the SR bias:
\begin{equation}\label{eq:sr_bias_long}
    \text{bias}\left(\widehat{\Theta}^{*}_p\right) = \tfrac{1}{2\theta}\tfrac{\sigma_\mu^2}{\sigma_R^2} \left(\tfrac{\sigma_\mu^2}{\sigma_R^2}A_{n,\phi}^2 + 2(\theta^2-1)A_{n,\phi} + (\theta^2+1)B_{n,\phi} \right)
\end{equation}

Deriving a simple, tight analytical bound from this complex equation is challenging, particularly due to the potential divergence near $\theta = 0$. Nonetheless, Equation \ref{eq:sr_bias_long} still highlights meaningful qualitative relationships between the bias of the IID resampled Sharpe Ratio and the core variables: the autocovariance structure, and the asset's Sharpe Ratio:
\begin{equation}\label{eq:bias_sr_bound_analytical}
    \text{bias}\left(\widehat{\Theta}^{*}_p\right) \leq \frac{1}{2}\frac{\psi}{\theta}(C-2)
\end{equation}
where $\psi$, $\theta$, and $C$ are defined as before. This bound seems to hold empirically, but is not informative in practice when $\theta \approx 0$, since the bound explodes. Since accurate estimation is particularly important when $\theta \approx 0$, we investigate an alternative bound which is less interpretable but avoids the divergence. 

By taking a Taylor Approximation of the IID estimated Sharpe Ratio around an alternative point in Equation \ref{eq:taylor} -- specifically, the portfolio rule's true Sharpe Ratio -- may avoid the divergence, but introduces terms that are difficult to simplify algebraically. Starting this way, we propose an alternative bound which we motivate numerically in Appendix \ref{appssec:num_bound}:
\begin{equation}\label{eq:bias_sr_bound_numerical}
    \text{bias}\left(\widehat{\Theta}^{*}_p\right) \leq \theta - \tfrac{\theta^2 + \psi}{\sqrt{\theta^2 + C\psi}}.
\end{equation}
Substituting the bounds for the mean (Proposition \ref{prop:bias-mean}) and variance (Proposition \ref{prop:bias-var}) into $\Theta_p$ yields this bound for the Sharpe Ratio. While this approach may appear logically invalid, the true Sharpe Ratio displays interesting non-linear behaviour with the rolling window size $n$, which we use to motivate this bound (see Appendix \ref{appssec:num_bound} for details). Additionally, the bound is nonlinear in the variables $\theta$ and $\psi$, and its behaviour is less intuitive. Therefore, we use the analytically motivated bound from Equation \ref{eq:bias_sr_bound_analytical} to explain the behaviour of the bias of the Sharpe Ratio, but we use the numerically motivated bound from Equation \ref{eq:bias_sr_bound_numerical} for practice. For all empirical work in Sections \ref{sec:simulation} and \ref{sec:historical}, we use the bound from \ref{eq:bias_sr_bound_numerical}. We compare it against the bound in Equation \ref{eq:bias_sr_bound_analytical} in Appendix \ref{appsec:emp_comparison}.





\section{MonteCarlo Simulation}\label{sec:simulation} 

Our key analytical results are that the autocovariance structure is the primary driver of the bias of the resampled backtest, acting through two components: within the training set and between the training and test set. Furthermore, the bias is bounded by a function of the first-lag autocorrelation. Our simulation section aims to quantify and verify the relationships and bounds established analytically, under more realistic conditions, relaxing some of the assumptions. 

Our MonteCarlo experiment uses a standard set-up to examine the bias for a cross-section of assets with different parameters. Namely, we specify parameters of our DGP (see Assumption \ref{ass3}) informed by historical data. We use this DGP to simulate sample paths, which are used to calculate the distribution of the statistics of interest --- in this case, the bias of the resampled backtest. 

\subsection{Data}\label{ssec:data}
We use JKP Factor data \cite{jensen2023there} to specify realistic parameters for our assumed DGP. The JKP factor data contains monthly returns from 153 zero-cost US factor portfolios. The factor portfolios are constructed from different firm characteristics, clustered into 13 themes. Factor returns exhibit higher autocorrelation than individual stocks \cite{gupta2019factor}, representing a case where bias may be large. We use data from July 1972 to December 2023 for our subsequent analyses, with no missing data. However, our DGP is not trivial to estimate directly, so we use an ad-hoc process to select parameters that are consistent with ranges reported empirically. Appendix \ref{appssec:parm_spec} describes our parameter specification process in further detail. Summary statistics for the JKP data (which are also used as our simulation parameters) are presented in Appendix \ref{appssec:eda}


\subsection{Methodology}
We generate $K=1000$ sample paths of $T=480$ timesteps for each asset in our simulation universe to calculate the bias across the cross-section. For each simulated sample path $k = 1, \dots, K$, we perform: 1) an ordinary backtest, and then 2) an IID-resampled backtest. This gives Sharpe Ratio estimates $\widehat{\Theta}_{p,k}$ and $\widehat{\Theta}^{*}_{p,k}$, respectively. 

The bias is approximated as the average difference across all $K$ sample paths:
$$
\text{bias}\left(\widehat{\Theta}^{*}_p\right) = \frac{1}{K}\sum_{k=1}^{1000} \left(\widehat{\Theta}^{*}_{p,k} - \widehat{\Theta}_{p,k}\right).
$$
We similarly calculate the bias of the resampled backtest estimators for the portfolio rule's mean and variance. 

We perform backtests for a rolling-window MV portfolio rule that allocates between a single risky and risk-free asset. We assume a window size of $n=60$ months and a zero risk-free rate.  We use rolling estimates for both mean and variance (relaxing Assumption \ref{ass1}), setting risk aversion to $\gamma = 100$.  

We also assess the impact of several extensions on the bias of the resampled backtest estimators. 

\begin{enumerate}
    \item \textbf{Conditional Heteroskedasticity}: To assess the impact of relaxing Assumption \ref{ass3} on the resampled backtest estimators, we simulate noise terms from a univariate GARCH (1,1) process for each asset:
    $$
    \sigma_{m,t}^2 = \omega_m + \alpha_m \epsilon_{m,t-1}^2 + \beta_m \sigma_{m,t-1}^2, \quad m = 1, \dots, 153.
    $$
    where $m$ denotes the index of the asset. Parameters are chosen to match empirical persistence while maintaining the same long-run variance as the constant variance case. 
    \item \textbf{Portfolio Dimension}: We investigate how the bias changes as we increase the number of assets $M$ in the portfolio. We simulate portfolios with increasing $M$, iteratively adding assets to the portfolio, isolating the dimensionality effect from changes in underlying parameters.
    \item \textbf{Resampling Blocksize}: To study the effect of partially preserving temporal structure, we perform block resampling (with replacement) using various block sizes ($b=1, 2, 5, 10, 20$), where $b=1$ corresponds to IID resampling, and calculate the corresponding bias $\text{bias}\left(\widehat{\Theta}^{*(b)}_{p,k}\right)$
\end{enumerate}

\subsection{Results}\label{ssec:results}

\begin{figure*}[h!]
    \centering
    \includegraphics[width=0.9\linewidth]{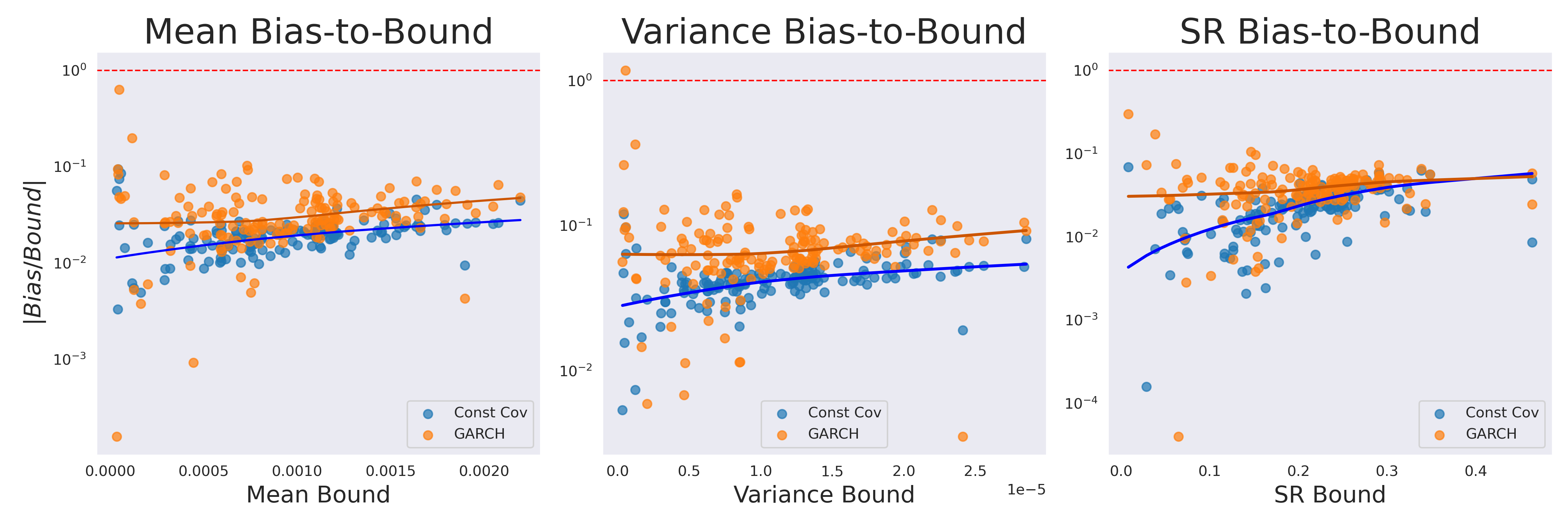}
    \caption{The simulated absolute value of the bias-to-bound ratios of the mean, variance, and Sharpe Ratio from an IID resampled backtest of a rolling window MV portfolio rule for different assets, plotted on a log scale. Each dot corresponds to a different asset. Linear (dashed) and Lowess (solid) trendlines are included to indicate the relationship with the bound. We compare the bias-to-bound ratios for two cases for the DGP: constant variance (blue) and GARCH (orange). The bias-to-bound ratio is typically very small, indicating the conservativeness and robustness of the bounds. GARCH effects increase the bias, but note that bias is plotted in a log-scale; on an absolute basis, the difference between the constant var and GARCH bias-to-bound ratios is negligible.. The bias-to-bound ratio is largest when the bound is approximately $0$, and can even fail in certain instances (as seen in the variance plot) because of effects not incorporated in the analytical bounds (here, the GARCH effects). }
    \label{fig:sim_bias_bounds}
\end{figure*}

We begin by numerically verifying the analytical bounds derived in Section \ref{sec:biasbounds}, and assess their robustness to conditional heteroskedasticity. Figure \ref{fig:sim_bias_bounds} plots the ratio of the simulated bias of the IID resampled backtest's mean, variance, and SR to the corresponding bound, for the aforementioned scenarios.

The results show that the bias-to-bound ratio consistently remains below $0.1$, verifying the bounds under conditional heteroskedasticity and indicating how conservative the bounds are. Although the ratio increases slightly with the bound, the absolute difference between the bias and bound increases with the bound (not shown) and this distance is approximately equal to the bound itself.  The bias-to-bound ratio increases under conditional heteroskedasticity for the mean and the variance, but the bias-to-bound ratio in the SR is largely the same as the constant variance scenario. This confirms our bounds provide reliable, albeit conservative, estimates for the potential maximum bias, even under a moderately more complex DGP. 

Having confirmed the bounds, we investigate what the primary drivers of the magnitude of the bias are by relating the bias to the underlying autocovariance structure. Our analytical framework links bias to first-order training-test dependence and the within-test dependence, reflected by the quantities:
\begin{enumerate}
\item{{\it Train-Test Dependence (TTD):}}
\begin{align}
    \mathrm{TTD} =\frac{1}{n}\sum_{i=1}^n \text{cov}\left(R_{t+1-i},R_{t+1}\right) = \sigma^2_\mu A_{n,\phi}
    \label{eq:tt-dependence}
\end{align}
\item{{\it Within-Training Dependence (WTD):}} 
\begin{align}
    \mathrm{WTD}= \frac{1}{n^2}\sum_{i=1}^n \sum_{j \neq i}^n \text{cov}\left(R_{t+1-i},R_{t+1-j}\right) = \sigma^2_\mu B_{n,\phi}.
    \label{eq:wt-dependence}
\end{align}
\end{enumerate}
Under Assumptions \ref{ass1} - \ref{ass3}, the bias is linear in the dependence components. However, our simulation results show that the TTD and WTD are almost scalar multiples of one another. Recall from Section \ref{ssec:biasbounds} that the sums $A_{n,\phi}$ and $B_{n,\phi}$ become increasingly linearly related as $n$ increases. Under our simulation DGP, it appears that $n=60$ is sufficient for approximate linearity between $A_{n,\phi}$ and $B_{n,\phi}$. Therefore, the relationship of TTD and WTD with the bias appears almost identical. To avoid duplicating plots, we only present the relationship between the bias and the TTD. 

\begin{figure*}[thb!]
    \centering     
    \includegraphics[width=0.9\linewidth]{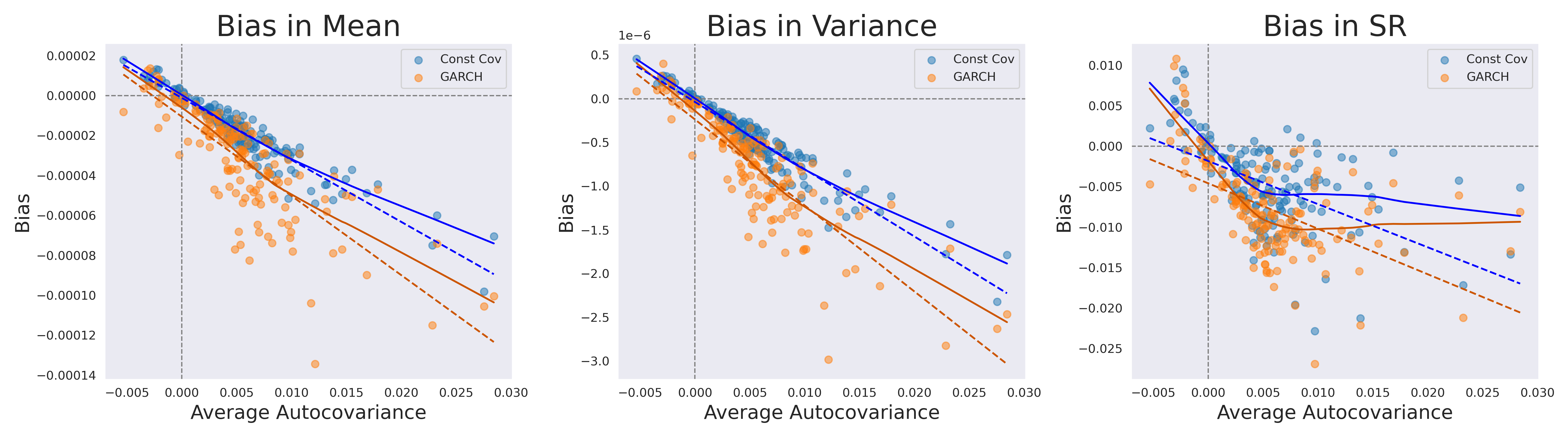}
    \caption{The drivers of bias of IID resampled backtests: the autcovariance structure. We plot the bias of the IID-resampled backtest estimators for rolling MV portfolios against the underlying asset's returns average autocovariance over the rolling window for the constant var (blue) and GARCH (orange) cases. Linear (dashed) and Lowess (solid) trendlines are included. The plots show that the bias in the mean (left) and the variance (middle) is largely linearly related to the average autocovariance for both the constant var and GARCH cases. The dispersion in bias increases with the average autocovariance, resulting in a nonlinear trendline as points with more extreme autocovariances have greater influence on the local average. The dispersion also increases with GARCH. The bias in the Sharpe Ratio (right) appears to be nonlinearly related to the average autocovariance, but this is misleading. The bias of the IID resampled Sharpe Ratio is actually linearly related to the autocovariance, where the gradient of the line is related to the inverse of the asset's Sharpe Ratio (see Figure \ref{fig:sim_sr})}
    \label{fig:sim_drivers}
\end{figure*}

\begin{figure}[bht!]
    \centering
    \includegraphics[width=\columnwidth]{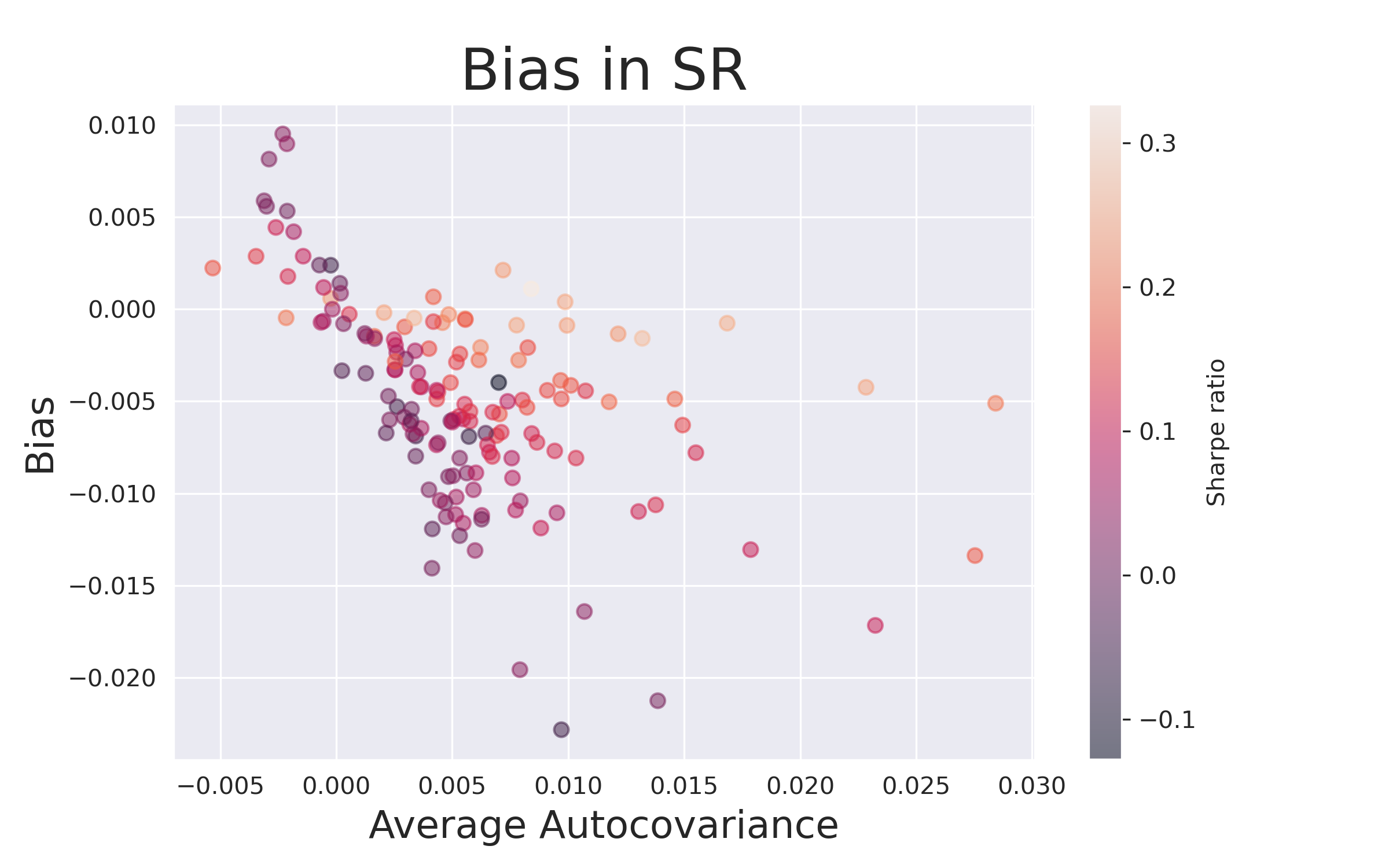}
    \caption{Bias of portfolio rule's Sharpe Ratio against average autovariance, coloured by the asset's Sharpe Ratio. The Sharpe Ratio appears to be nonlinearly related to the autocovariance in Figure \ref{fig:sim_drivers}. Here, we see that the portfolio rule's Sharpe Ratio appears to be linearly related to the average autocovariance, but where the gradient of the line depends on the inverse of the asset's Sharpe Ratio. IID resampled Sharpe Ratios for the MV portfolio are most affected by disruption in the temporal dependence when the underlying asset's Sharpe Ratio is small. }
    \label{fig:sim_sr}
\end{figure}

Figure \ref{fig:sim_drivers} plots the bias against the TTD when assumptions are relaxed (i.e., estimated variance, GARCH errors). The biases in the mean and variance are negative and strongly linearly related to the TTD and WTD (not shown), with $R^2 \approx 0.9$. With the inclusion of GARCH effects, the linear relationship between the biases in the mean and variance is weaker, with the $R^2$ dropping to $0.63$ and $0.73$, respectively. Nonetheless, the first-order covariance through the TTD and WTD explains the majority of cross-sectional variation in bias. Figure \ref{fig:sim_drivers} shows that the Bias in the Sharpe Ratio appears to have a negative relationship with the dependence. Yet, we see that the relationship no longer appears to be clearly linear, with an $R^2 \approx 0.28$. Closer inspection of the plot shows that the bias variation increases as the autocovariance moves away from zero. The observed variation masks a more complicated, albeit linear, relationship between bias and dependence, where the underlying asset's Sharpe Ratio governs the relationship. This is shown more clearly in Figure \ref{fig:sim_sr}, where each asset's bias is coloured according to its asset's Sharpe Ratio, showing that the gradient of the linear relationship is inversely related to the asset's Sharpe Ratio.



\begin{table}[thb!]
\centering
\caption{Goodness-of-fit ($R^2$) from regressing bias against average autocovariance for the constant variance (Const. Cov.) and GARCH cases. The $R^2$ for the bias in the mean and variance decreases with the addition of GARCH effects. The $R^2$ for the bias of the Sharpe Ratio is much smaller than the bias of the mean and the variance, reflecting its dependence on the asset's Sharpe Ratio as well.}\label{tab:sim_r_square}
\begin{tabular}{lrrr}
\toprule
{} &  bias($\widehat{\mu}^{*}_p$) &  bias($\widehat{\sigma}^{*2}_p$) &  bias($\widehat{\Theta}^{*}_p$) \\
\midrule
Const. Cov.  &       90\% &          91\% &       28\% \\
GARCH &       63\% &          73\% &       28\% \\
\bottomrule
\end{tabular}
\end{table}

\begin{table*}[thb!]
\centering
\caption{Magnitude of bias. We examine the cross-sectional distribution of the absolute value of the bias relative to the standard errors of the original backtest estimators, for rolling MV portfolio rules allocating between a single risky and risk-free asset. The IID Resampled Sharpe Ratio has a lower standardised bias than the IID resampled mean and variance. The standardised bias increases as GARCH effects are incorporated. Nonetheless, the standardised bias of the IID resampled Sharpe Ratio may be considered small, with the 95th percentile $0.33$. Assuming a Normal distribution for the standard backtest estimator, a standardised backtest produces estimates outside of $0.33$ standard errors from its expected value with a probability of nearly $0.8$. However, results here are for $M=1$, the standardised bias increases with dimension (see Figure \ref{fig:sim_dimension})  }\label{tab:sim_mag_bias_se}
\begin{tabular}{crrrrrr}
\toprule
{Percentile} & \multicolumn{3}{c}{Constant Covariance} & \multicolumn{3}{c}{GARCH} \\
{} & $\frac{\text{bias}(\widehat{\mu}^{*}_p)}{\text{se}(\widehat{\mu}_p)}$ & $\frac{\text{bias}(\widehat{\sigma}^{*2}_p)}{\text{se}(\widehat{\sigma}^{2}_p)}$ & $\frac{\text{bias}(\widehat{\Theta}^{*}_p)}{\text{se}(\widehat{\Theta}_p)}$ & $\frac{\text{bias}(\widehat{\mu}^{*}_p)}{\text{se}(\widehat{\mu}_p)}$ & $\frac{\text{bias}(\widehat{\sigma}^{*2}_p)}{\text{se}(\widehat{\sigma}^{2}_p)}$ & $\frac{\text{bias}(\widehat{\Theta}^{*}_p)}{\text{se}(\widehat{\Theta}_p)}$ \\
\midrule
0.05 &     0.03 &        0.05 &     0.01 &     0.03 &        0.05 &     0.02 \\
0.25 &     0.09 &        0.19 &     0.04 &     0.15 &        0.30 &     0.09 \\
0.50 &     0.15 &        0.30 &     0.09 &     0.22 &        0.43 &     0.16 \\
0.75 &     0.20 &        0.40 &     0.15 &     0.28 &        0.53 &     0.21 \\
0.95 &     0.30 &        0.60 &     0.26 &     0.36 &        0.65 &     0.33 \\
\bottomrule
\end{tabular}
\end{table*}

Having examined the drivers, we now assess the practical magnitude of the bias introduced by IID resampling. To provide context beyond absolute bias values (which depend on scale), we compare the estimated bias of the Sharpe Ratio estimator to the estimation uncertainty inherent in the standard backtest procedure. Specifically, we calculate the following ratio by simulation:
$$
    \text{Magnitude Ratio} = \frac{\text{bias}\left(\widehat{\Theta}^{*}_p\right)}{\text{SE}\left(\widehat{\Theta}_p\right)}.
$$
where $\text{std err}\left(\widehat{\Theta}_p\right)$ is the standard error of the standard backtest estimate. Table \ref{tab:sim_mag_bias_se} summarises the cross-sectional distribution of the magnitude ratio observed in our simulations.  The standardised biases are largest for the variance and smallest for the Sharpe Ratio. Furthermore, the standardised biases for the mean, variance, and Sharpe Ratio all increase by including GARCH effects. The bias from IIS resampling appears to be modest relative to the standard error of the standard backtest, with a 95th percentile of the standardised biases of $0.33$ for the Sharpe Ratio. 

\begin{figure}[thb!]
    \centering     
    \includegraphics[width=\linewidth]{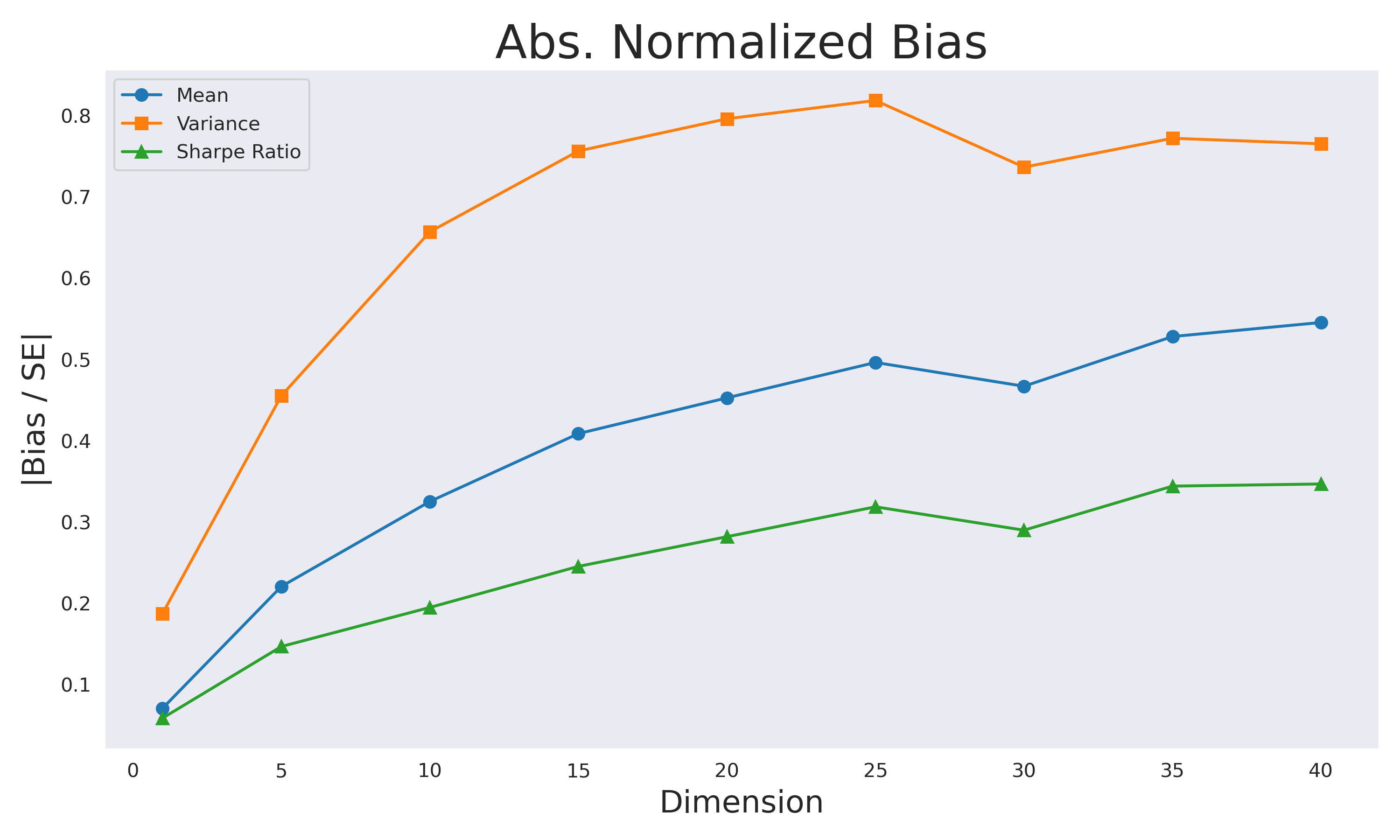}
    \caption{Bias by Dimension. Left: the growth of the bias with dimension plotted on a log scale. Center: the growth of standard error plotted on a log scale. Right: the bias normalised by the standard error against dimension. The bias normalised by the standard error grows at a slowing rate for the mean, variance, and SR.}
    \label{fig:sim_dimension}
\end{figure}

Table \ref{tab:sim_mag_bias_se} presents the case for $M=1$. The standardised bias grows with the dimension as seen in Figure \ref{fig:sim_dimension}, where the standardised bias of the Sharpe Ratio for $M=40$ is approximately three times as large as $M=1$. The standardised bias appears to grow at a slowing rate, and it may reach an upper limit as $M$ gets increasingly large. We defer deeper analysis into the dimension effects for further study.

\begin{figure}[thb!]
    \centering
    \includegraphics[width=\linewidth]
    {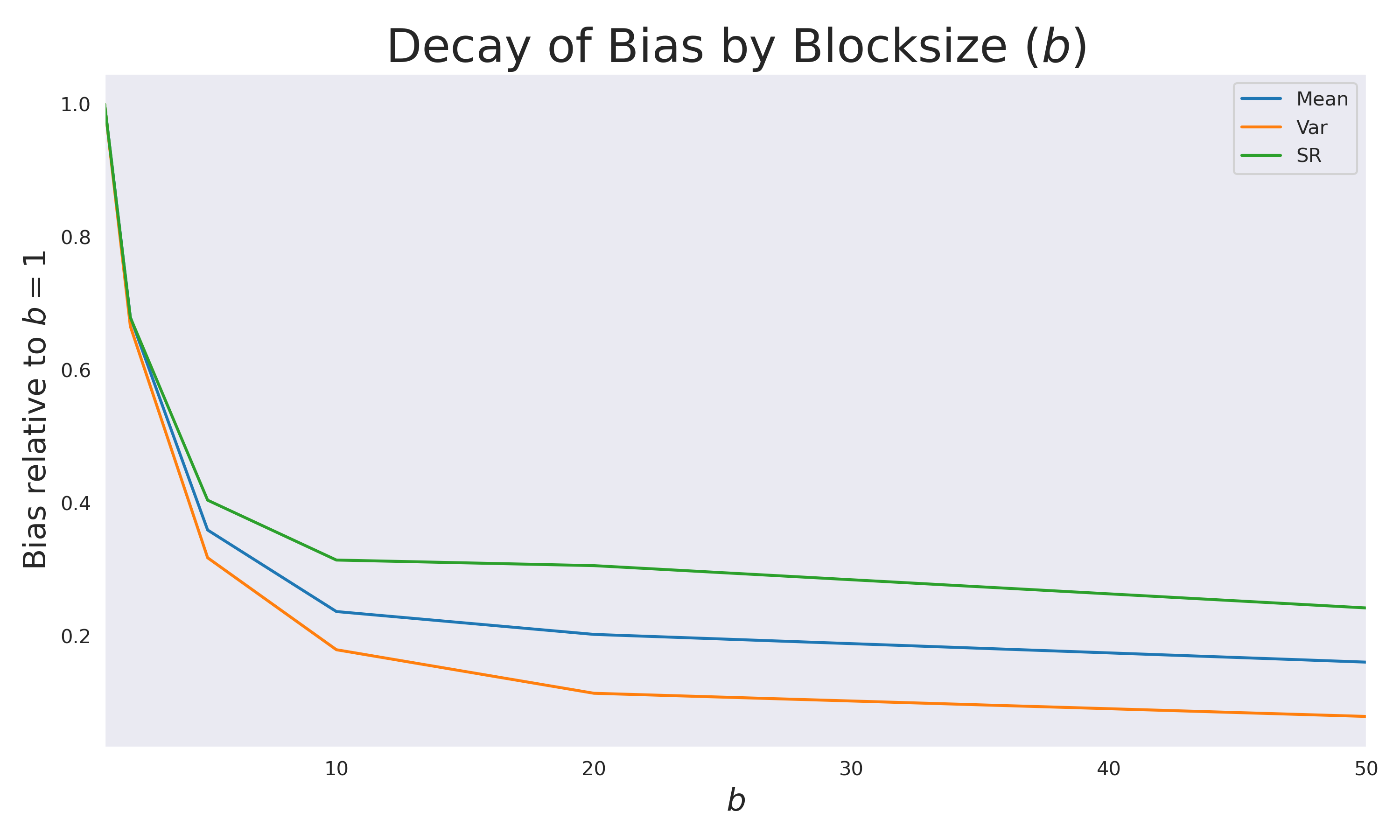}
    \caption{Bias by Blocksize. We plot how the bias decays (relative to IID resampling ($b=1$) with increasing block size. Most of the benefit from increasing block size occurs at $b=10$, and plateaus thereafter.}
    \label{fig:sim_bias_by_blocksize}
\end{figure}

\begin{figure}[thb!]
\centering    
    \includegraphics[width=\columnwidth]{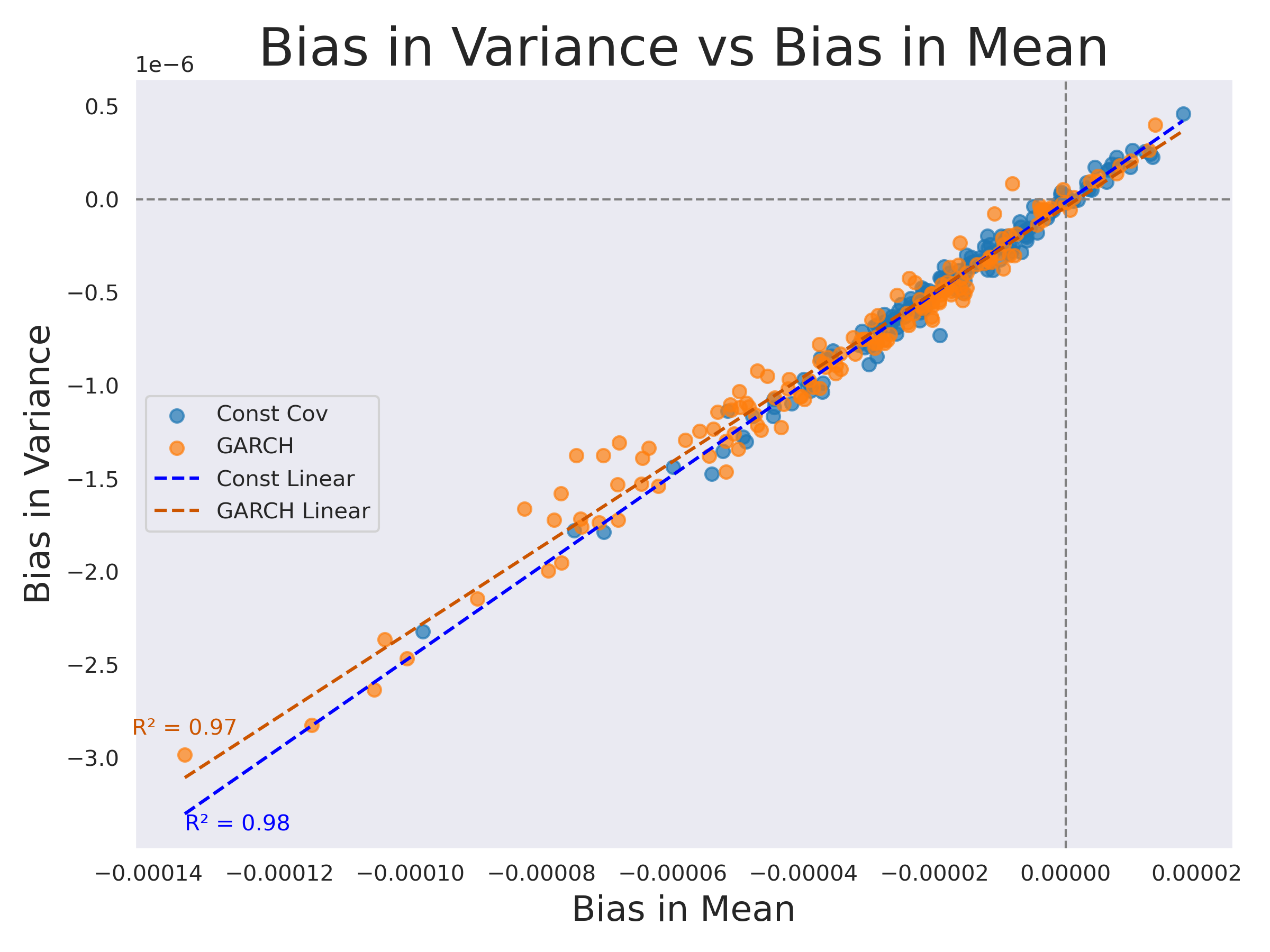}
    \caption{Relationship between bias in mean and variance. The plot shows a nearly linear relationship between the bias in the mean and the variance for both the constant variance and GARCH cases. Since the bias in the SR offsets the bias in the mean and the variance, this relationship leads to lower standardised biases in the SR.}
    \label{fig:sim_relationship}
\end{figure}

Table \ref{tab:sim_mag_bias_se} shows that the Sharpe Ratio has the lowest standardised bias. Furthermore, Figure \ref{fig:sim_dimension} shows that the bias in the Sharpe Ratio grows slowest. Both observations reflect the offsetting effect of the bias described earlier in Section \ref{ssec:biasbounds}. Figure \ref{fig:sim_relationship} demonstrates a near-perfect linear relationship between the bias in the mean and the bias in the variance, which facilitates the offsetting mechanism even with GARCH effects. While this offsetting is inherent, block resampling offers an external mitigation approach. Figure \ref{fig:sim_bias_by_blocksize} suggests that partially preserving temporal can significantly reduce bias. The bias decreases rapidly, with an approximately $60\%$ decrease as the block size $b$ increases from 1 (IID) to $b=10$, but with diminishing marginal benefits as $b$ increases.

\subsection{Summary}
Our simulation study provides strong numerical support for the analytical findings under less strict conditions. The analytical bounds were confirmed to hold conservatively, even in the presence of GARCH effects. The analysis of bias drivers demonstrated that the disruption of the autocovariance structure drives the bias primarily through the disruption of train-test dependence, linked to the sum $A_{n,\phi}$. Quantitatively, the bias of the Sharpe Ratio is modest for $M=1$, but increases with dimension. Further analysis is required to determine the limiting relationship between bias and dimension. From a mitigation perspective, the Sharpe Ratio has an inherent mitigation mechanism that offsets component biases, which are linearly related. Block resampling can be used to further mitigate bias.

\section{Historical Data}\label{sec:historical}

\begin{figure*}[thb!]
    \centering
    \includegraphics[width=\textwidth]{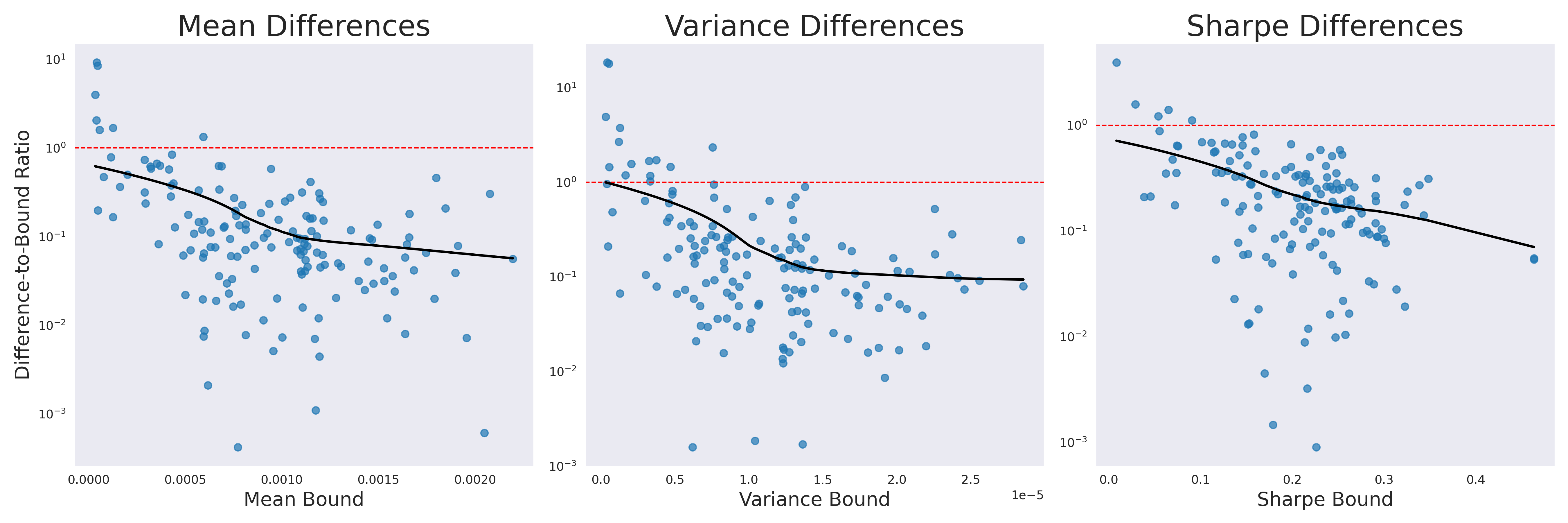}
    \caption{The absolute value of the ratio of differences to the estimated bounds plotted on a log-scale. The red line represents the threshold where the ratio equals 1, showing that at least $90\%$ of differences are below the threshold for each plot. The black line is a LOWESS trend line to indicate the presence of any trend that may exist, and appears to be in the opposite direction from the simulation results. As expected, the empirical differences are more dispersed than the simulated biases. The bound is violated when close to zero (i.e., when autocorrelation is very low).   }
    \label{fig:empirical_bounds}
\end{figure*}

Empirical data has many features (so-called ``stylised facts" \cite{cont2001empirical}) that are typically not considered, at least not altogether, in models for asset returns. For example, empirical returns also display negative skewness, heavy tails, mean reversion, and regime changes. This section examines the behaviour of IID-resampled backtests on historical data (JKP Factors, July 1972 to Dec 2023 --- see Section \ref{ssec:data}) to assess consistency with our prior findings and explore real-world applications. In addition, we draw attention to the long-memory effect \cite{WilcoxGebbie2008} and its impact on autocorrelations.

Ideally, we want to see if our analytical findings are true under real-world processes. Given that we only observe a single sample path, we can't directly estimate bias without making strong assumptions about the structure of the underlying process. Instead, our empirical analysis investigates the cross-sectional distribution of differences between the IID resampled backtests and the standard backtests for each asset:
\begin{equation}\label{eq:diff}
    d_m = \widehat{\Theta}^{*}_{p,m} - \widehat{\Theta}_{p,m} \ , \quad m=1, \dots, 153
\end{equation}
where $m$ denotes the index of each asset. A cross-sectional analysis of the differences partially mitigates the sampling error inherent in the individual differences, allowing us to compare whether general patterns are consistent with our previous findings. 
We investigate the differences against empirical estimates of the analytical bounds derived in Section \ref{sec:biasbounds}. The bounds are estimated on the original (as opposed to resampled) data using sample moment estimates instead of parameters. Figure \ref{fig:empirical_bounds} plots the ratio of the differences to estimated bounds against the estimated bounds. The bounds appear to still hold (ratios are less than $1$) even with sampling errors, highlighting the conservative nature of the bounds. Contrary to our simulation results, we see that many differences are positive. Furthermore, while our simulation results showed that the bias-to-bound ratio decreased with the bound (typically as the lag-1 autocorrelation increases), our empirical results show no clear relationship between the differences and the bounds itself. While sampling error obscures patterns, we attribute both the positive differences and the lack of relationship between the differences and bounds to the deviations between the empirical and assumed autocovariance structures, which we discuss further. 

\begin{figure}[bht!]
    \centering
    \includegraphics[width=\columnwidth]{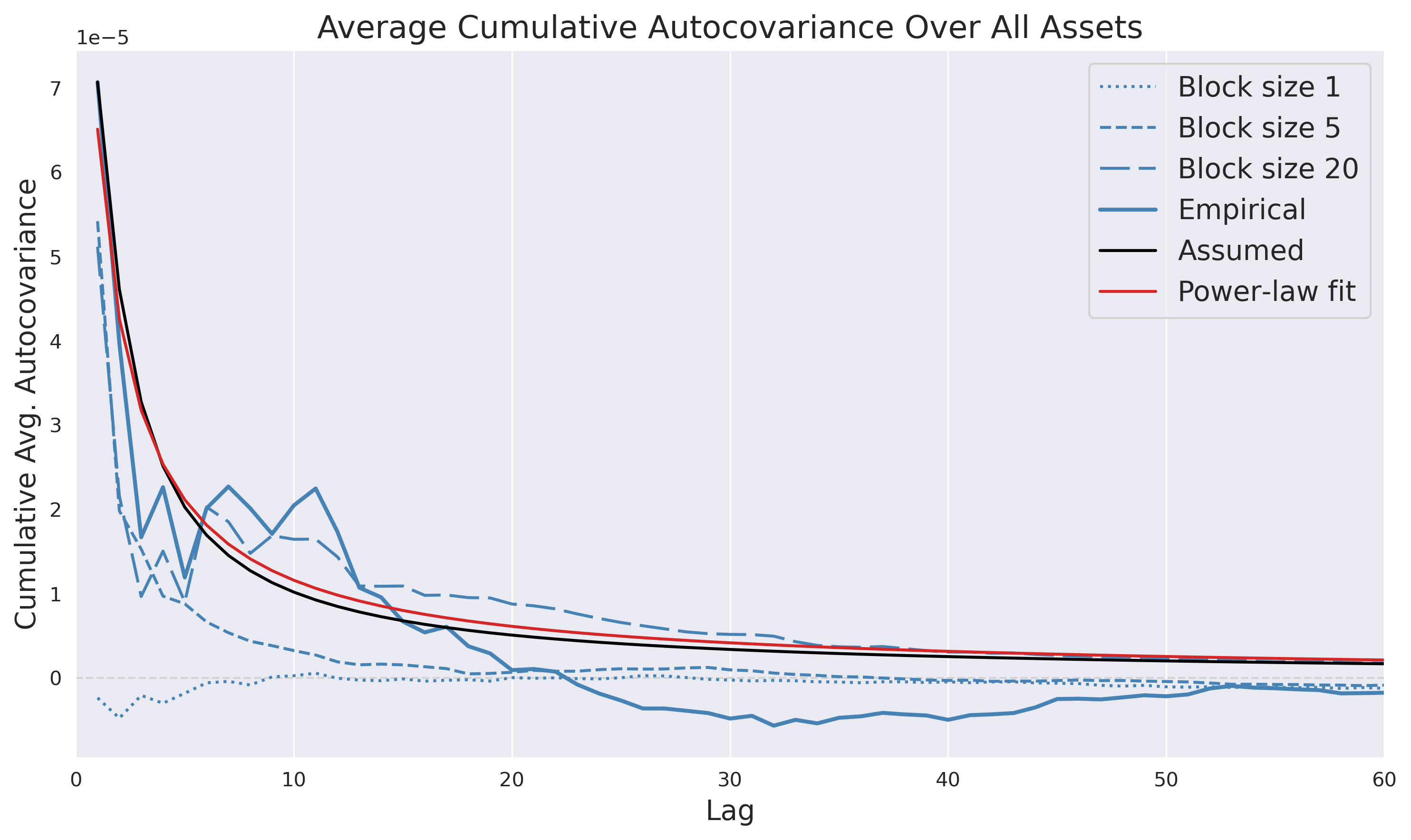}
    \caption{A comparison of autocovariance structures. We calculate the cumulative average autocovariance at each lag for the risky returns and plot the cross-sectional average for the original data, the resampled data, and the simulation DGP. The cumulative average autocovariance reflects the dependence between the in-sample and out-of-sample returns and is a key component of the bias from IID-resampled backtests. Whereas the simulated autocovariance is always positive, the negative cumulative empirical autocovariance at lag 60 (the window size used in our results) produces differences between the IID and resampled backtests with the opposite sign to that assumed. Furthermore, while the simulated curve decays slowly to zero, the empirical curve shows that the average autocovariance stays constant up to about a year, and then decays becomes negative, and then reverts to zero. IID Resampling erodes the autocovariance structure, but block resampling partially preserves the structure. As block size increases, the rate of decay decreases. }
    \label{fig:empirical_autocovariance}
\end{figure}

\begin{figure*}[thb!]
    \centering
    \includegraphics[width=\textwidth]{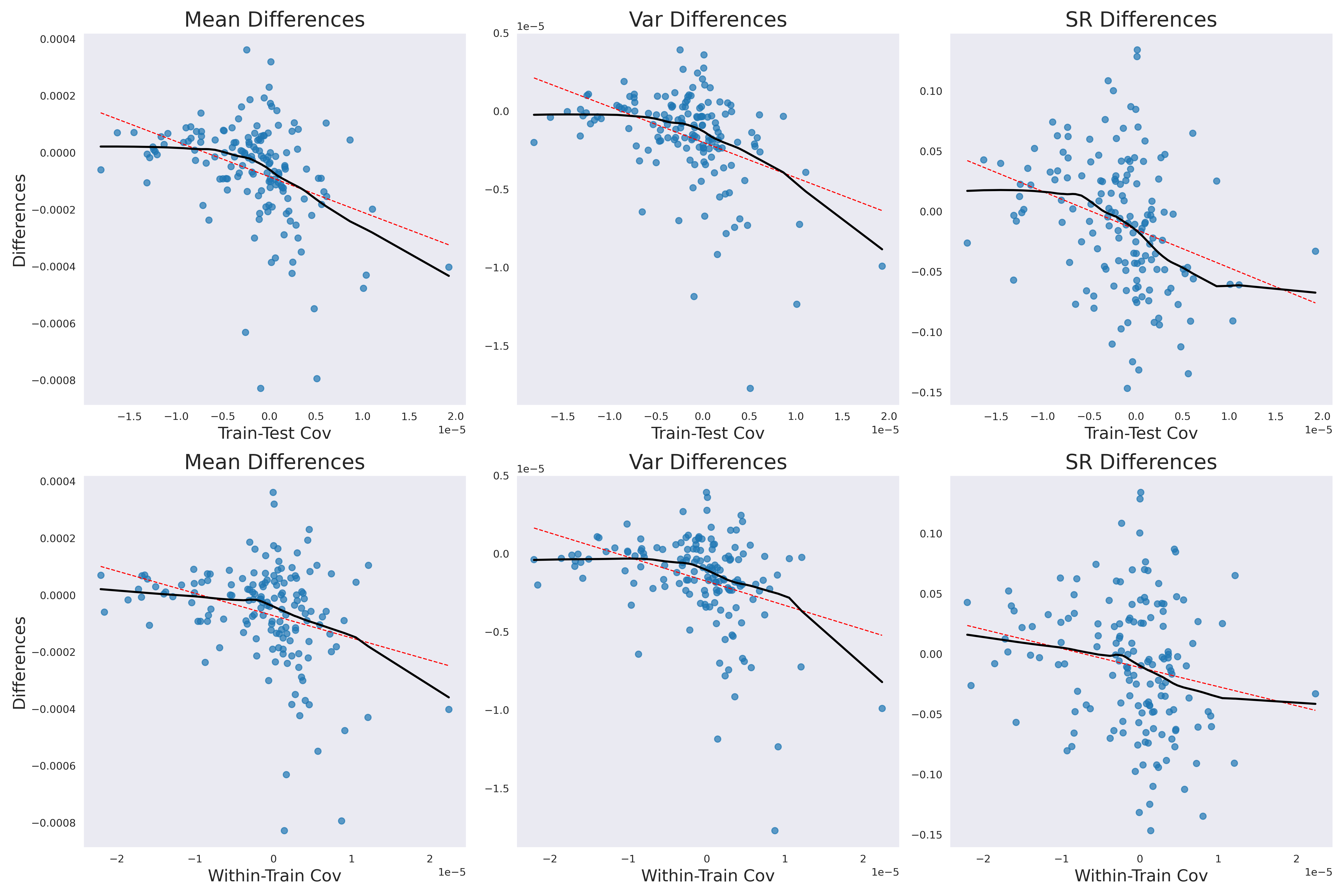}
    \caption{Differences against estimated dependence structures. A Linear Trend is fitted in red, and a LOWESS trend is fitted in solid black. All plots show negative relationships with the dependence components. The relationship is weak, but statistically significant (also see Table \ref{tab:emp_rsquare}). It is not clear if the nonlinear trend shown by the lowess smoothing indicates underlying nonlinearity or reflects noisy estimates, particularly at the edges of the plot.   }
    \label{fig:empirical_drivers}
\end{figure*}

To investigate the drivers, we examine the relationship of the estimated differences with the autocovariance structure. In Figure \ref{fig:empirical_drivers}, we plot the differences against the estimates of the between-training and within-training dependence components, identified in Equation \ref{eq:diff}. To estimate the dependence components, we first estimate the autocovariance function using each asset's sample path, from which we directly calculate the sums:
$$
\begin{aligned}\label{eq:emp_dependence}
&\frac{1}{n}\sum_{i=1}^n \widehat{\text{cov}}\left(R_{t+1-i}, R_{t+1} \right) \\
&\frac{1}{n}\sum_{i=1}^n \sum_{j\neq i}^n \widehat{\text{cov}}\left(R_{t+1-i}, R_{t+1-j} \right)
\end{aligned}
$$
where $n$ is the window length. Table \ref{tab:emp_rsquare} presents the $R^2$ values from regressing the differences against the estimated dependence components. In comparison with our simulation results, Figure \ref{fig:empirical_drivers} shows the differences $d_m$ have greater dispersion, which is to be expected because of the sampling error. Consequently, the $R^2$ values are much lower than for our simulation results (see Table \ref{tab:sim_r_square} for comparison), but the $R^2$s are all statistically significant with $p$-values $\approx0$, demonstrating that dependence components are still significant drivers. 


Notably, we observe that positive $d_m$ in the means and SRs corresponds to negative estimates of train-test dependence. These negative estimates indicate deviations between our assumed and the empirical autocovariance structure, which is visualised in Figure \ref{fig:empirical_autocovariance}. Our assumed autocovariance structure has a large positive autocovariance at the first lag that propagates to higher lags, weakening over lags. Empirical estimates show that autocovariance is positive and largest at the first lag, explaining why the bounds still hold empirically. However, unlike the assumed DGP, the autocovariance at higher lags does not appear to be a function of the first lag, explaining why there is no clear relationship between the differences and the bound, which depends primarily on the lag-1 autocorrelation. Nonetheless, the relationship between positive differences $d_m$ with the negative estimated train-test dependence is still consistent with our theoretical result $\text{bias}(\widehat{\mu}^{*}_p) = -\text{cov}(w_t, R_{t+1})$.

The absence of an empirical relationship between the differences in the variances and the estimated dependence components may reflect an underlying reality, or simply the relative magnitude of noise. We investigate whether cross-sectional differences in variances can be explained by conditional heteroskedasticity. Interestingly, the differences in variances also appear unrelated to the autocovariance of the squared residuals (not shown). This can be important in better understanding the presence of periodicity in the data {\it i.e.} calendar affects ({\it e.g.} Figure 2 in \citet{WilcoxGebbie2008}), as evidenced by the peaks seen in empirical cumulative autocovariances in Figure \ref{fig:empirical_autocovariance} between 6 to 9 months and then near 12 months. These are smoothed out by the block and IID resampling, as opposed to the long-memory effects, which appear not to be smoothed out to the same extent -- particularly when using block resampling as opposed to IID resampling. 

\begin{table}[htbp]
  \centering
  \caption{$R^2$ from regressing performance differences onto dependence components (all values in \%). Although the relationship is substantially weaker than the simulations, because of the estimation noise, all $R^2$ values are nonetheless significant at the $0.5\%$ level.}
  \label{tab:emp_rsquare}
  \begin{tabular*}{\columnwidth}{@{\extracolsep{\fill}} l r r r}
    \toprule
    & \multicolumn{3}{c}{\textbf{$R^2$ (\%)}} \\
    \cmidrule(lr){2-4}
    \textbf{Component}  & \textbf{Mean} & \textbf{Var} & \textbf{Sharpe} \\
    \midrule
    Train–Test          & $15$ & $17$ & $10$ \\
    Within–Train        & $9$  & $13$ & $4$  \\
    Both                & $16$ & $18$ & $15$ \\
    \bottomrule
  \end{tabular*}
\end{table}

Next, we investigate the magnitude of the differences. Table \ref{tab:empirical_tests} presents the distribution of studentised differences where the standard errors are estimated using a Stationary Bootstrap. We also test whether the differences are different from zero, indicating whether the differences are consistent with sampling variation, using the Studentised Bootstrap approach of \cite{ledoit2008robust}. We report the proportion of assets with rejected tests at the $5\%$ level. However, since the backtests are noisy, the test will have low power, making it difficult to detect small biases. Therefore, we also compare the magnitude of the differences with the standard error of the standard backtest for further context, test whether the differences lie within the confidence interval of the standard backtest estimates, and report the rejection rates. This indicates whether a standard backtest could reasonably produce IID resampled backtest estimates. 


\begin{table}[thb!]
  \centering
  \caption{Magnitude of Differences Relative to Standard Errors. The top section reports percentiles of the absolute value of studentized test statistics, where standard errors are computed for the difference between estimators. The bottom section reports the same percentiles using standard errors from the standard backtest alone. The final row in each section shows the proportion of test rejections. The studentised differences for the Sharpe Ratio are lower than for the mean and variance, highlighting the offsetting effect. At the $5\%$ level, most differences are statistically indistinguishable from $0$. However, the percentiles are likely to increase as we increase dimension.  }
  \label{tab:empirical_tests}
  \begin{tabular*}{\columnwidth}{@{\extracolsep{\fill}} l r r r}
    \toprule
    & \multicolumn{3}{c}{\textbf{Studentised Differences}} \\
    \cmidrule(lr){2-4}
    \textbf{Percentile / Test}
      & \textbf{Mean}
      & \textbf{Variance}
      & \textbf{Sharpe} \\
    \midrule

    \addlinespace[0.75ex]
    $0.05$       & $0.06$ & $0.12$ & $0.04$ \\
    $0.25$       & $0.33$ & $0.42$ & $0.30$ \\
    $0.5$        & $0.73$ & $0.98$ & $0.67$ \\
    $0.75$       & $1.21$ & $1.52$ & $1.10$ \\
    $0.95$       & $1.91$ & $2.53$ & $1.85$ \\

    \addlinespace[0.75ex]
    prop $p$-value $<0.05$ & $0.06$ & $0.12$ & $0.03$ \\

    \midrule

    \addlinespace[0.75ex]
    $0.05$            & $0.08$ & $0.20$ & $0.05$ \\
    $0.25$            & $0.46$ & $0.62$ & $0.44$ \\
    $0.5$             & $0.97$ & $1.31$ & $0.94$ \\
    $0.75$            & $1.56$ & $1.93$ & $1.48$ \\
    $0.95$            & $2.36$ & $3.05$ & $2.40$ \\

    \addlinespace[0.75ex]
    prop outside $0.95$ CI & $0.16$ & $0.35$ & $0.08$ \\

    \bottomrule
  \end{tabular*}
\end{table}

\begin{figure}[thb!]
    \centering
    \includegraphics[width=\columnwidth]{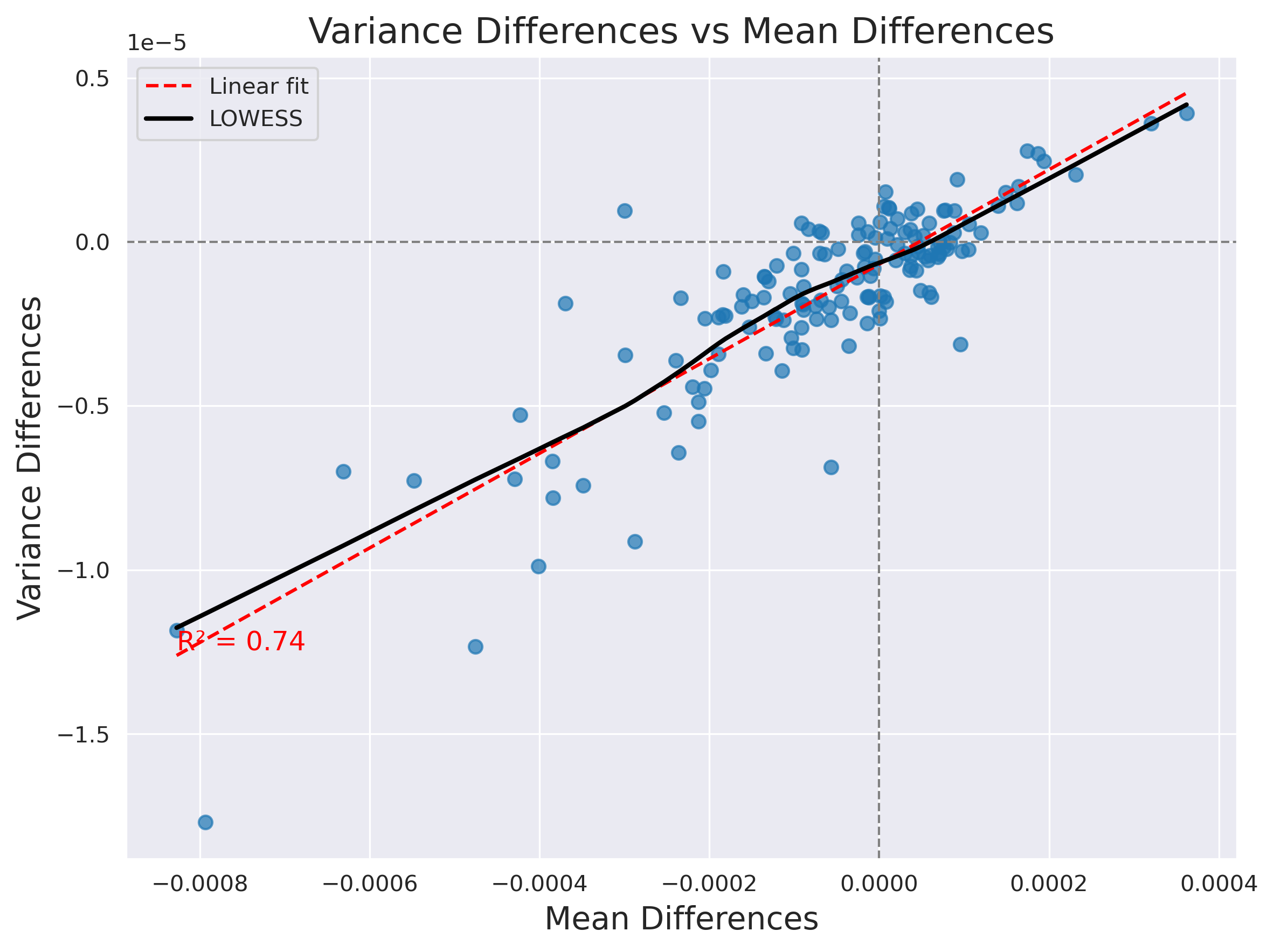}
    \caption{Relationship between differences in means and differences in variances. A Linear Trend is fitted in red, and a LOWESS trend is fitted in solid black. The $R^2=0.74$ indicates a strong linear relationship, verifying the relationship between the bias in the means and variances identified analytically and through simulation.}
    \label{fig:empirical_relationship}
\end{figure}

Our analytical and simulation results explained that the smaller bias of the SR is due to offsetting the biases in the mean and the variance. In Figure \ref{fig:empirical_relationship}, we examine whether the empirical differences $d_m$ also show a linear relationship between the means and the variances. 

As discussed earlier, we investigate how the differences change with dimension and the effects of block resampling. 

\begin{figure}[thb!]

    \begin{subfigure}[b]{\linewidth}
        \includegraphics[width=\linewidth]{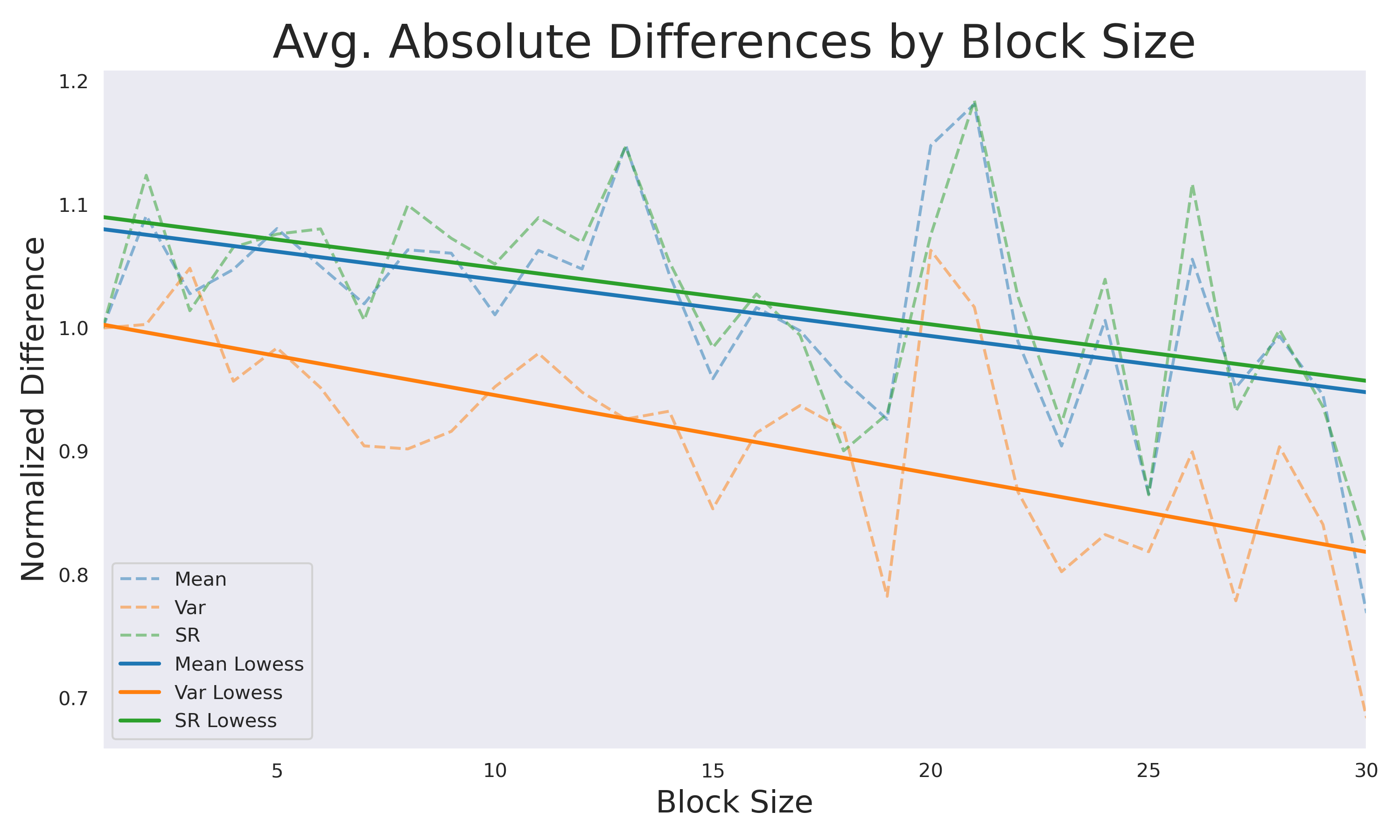}
        \caption{Decay in Abs. Value of Differences by Block size. We plot the cross-sectional average of the absolute value of the differences against blocksize, relative to a block size of 1 (dashed). Since estimates are noisy, we plot linear trendlines (solid) to highlight the decreasing trend. Increasing the blocksize from 1 to 15 reduces the bias by approximately $10\%$ for the mean, variance, and SR. }
        \label{fig:bias_by_blocksize}
    \end{subfigure}
    
    \begin{subfigure}[b]{\linewidth}
        \includegraphics[width=\linewidth]{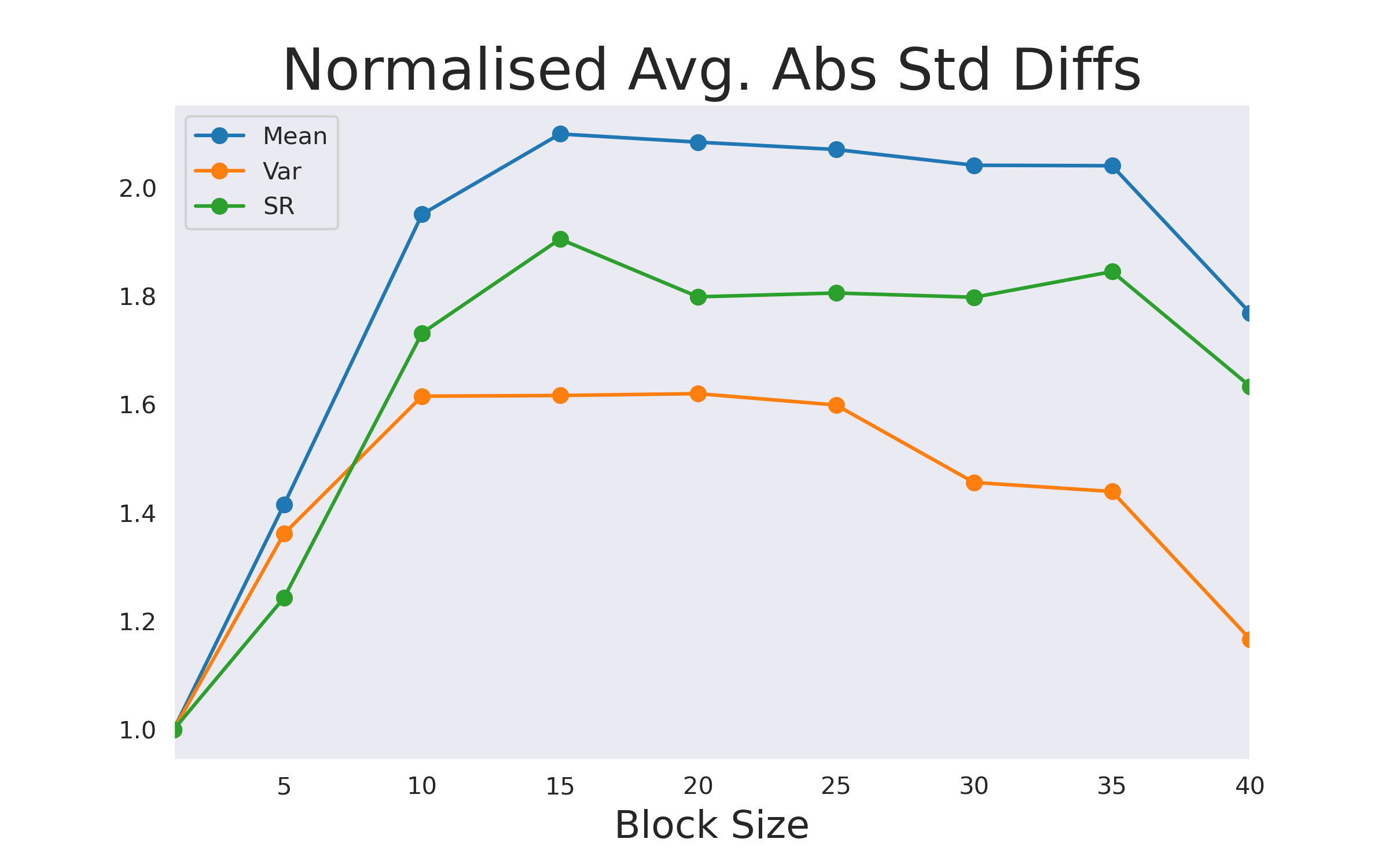}
        \caption{Growth in Standardised Differences by Dimension. We standardise the differences with the standard errors of the original backtest. We plot the cross-sectional average of the absolute value of the standardised differences, relative to dimension $M=1$. The standardised differences increase quickly and appear to plateau. Interestingly, the magnitude appears to decrease at higher dimensions, but this may be due to sampling variation.}
        \label{fig:bias_by_unknown_var}
    \end{subfigure}    

    \caption{Further Exploration of Bias, investigating the i) effect of blocksize, iii) dimension, iv) relationship between bias in mean and variance. }\label{fig:ancil_sim_results}

\end{figure}

\section{Discussion}\label{sec:discussion}

\subsection{Synthesis}
Standard backtests often provide unreliable performance estimates due to the limitations of relying on a single historical data path. Resampling methods aim to mitigate this by increasing the amount of data available. However, structure-agnostic approaches, like IID resampling, face a critical challenge of introducing estimation bias through the disruption of temporal dependence. 

Our investigation reveals that the primary driver of this bias is fundamentally the autocovariance structure of the underlying asset returns over the rolling window used by the portfolio rule. The rolling window size $n$ is critical in determining this relationship. 
While empirical asset returns often exhibit low first-lag autocorrelation $\psi$, their underlying moments (like risk premia or volatility) can possess strong persistence. 

\emph{Does this underlying moment persistence necessitate preserving time order?} Our analysis suggests no. The crucial factor for IID bias is the dependence structure present in the returns themselves. If strong moment persistence does not translate into significant return autocovariance ({\it e.g.} due to low Signal-to-Noise Ratio (SNR), as discussed in our analytical model), then the bias from disrupting time ordering via IID resampling will consequently be small. It is the disruption of the return autocovariance, not necessarily deeper moment dependencies, that fundamentally creates the bias.

Our analytical toy model allows us to isolate the key mechanism by simplifying reality. It demonstrates how signal persistence $\phi$ and signal strength (as measured by the SNR) combine to produce return autocovariance $\psi$, and hence drive the bias. Our analytical model incorporates temporal structure in the risk premium only. Our simulation results suggest that conditional heteroskedasticity (a GARCH-like structure) plays a secondary role compared to risk premium variation in determining SR bias. This most likely reflects the relative importance of mean {\it vs.} variance estimation for mean-variance (MV) decision based performance. 

Comparing the model's simple, monotonically decaying autocovariance assumption with empirical observations (Figure \ref{fig:empirical_autocovariance}) reveals key differences, in particular the presence of negative autocovariances at longer lags in real data. This empirical structure explains important observed phenomena, such as the positive differences $d_m$ found for some assets (Figure \ref{fig:empirical_drivers}), which occur when the average train-test covariance is negative. The empirical results still remain consistent with the mechanism that the autocovariance over the rolling window drives the bias.

Analytically, the disruption of the autocovariance structure manifests as bias through two primary channels: First, altering train-test dependence (TTD, related to terms like $A_{n,\phi}$). Second, the within-train dependence (WTD, related to terms like $B_{n,\phi}$). The TTD directly impacts the covariance between portfolio weights and future returns (Equation \ref{eq:biasm}) by affecting bias in the IID-resampled mean. While the WTD influences moments calculated solely within the training window and contribute more to bias in the IID-resampled variance. 

The estimation window size $n$ controls the relationship between these two components. Figure \ref{fig:sim_relationship} show that $A_{n,\phi}$ and $B_{n,\phi}$, and thus $\text{bias}(\widehat{\mu}^{*}_{p})$ and  $\text{bias}(\widehat{\sigma}^{*2}_{p})$, become increasingly linear as $n$ increases. This linearity clarifies the offsetting mechanism within the SR bias and simplifies its relationship with the underlying autocovariance structure; especially for larger window sizes.

Equation \ref{eq:bias_sharpe} uses a Taylor approximation to reveal that the SR bias results from a partial cancellation between the relative biases in the mean and variance. Figure \ref{fig:sim_relationship} then demonstrates this offsetting mechanism, and is supported empirically by the near-linear relationship between component biases. This often makes the backtested SR less sensitive to IID resampling than its components -- {\it i.e.} the backtest mean and variance. To assess practical significance, we compare the bias magnitude to the standard error of the original backtest. Table \ref{tab:sim_mag_bias_se} summarises this simulation work and suggests the standardised SR bias can be modest {\it e.g.} median $0.16$, and $0.33$ at the 95$^{th}$ percentile for $M=1$. This is especially pertinent when compared to the standardised biases for mean and variance. 

However, empirical differences ($d_m$) relative to their standard errors show greater variability, with a non-trivial fraction, potentially indicating larger relative differences. This is shown in Table \ref{tab:empirical_tests}; although statistical significance is often lacking due to noise and low test power. Furthermore, the standardized bias increases notably with portfolio dimension (Figure \ref{fig:sim_dimension}), though growth in the SR bias appears to slow down and plateau with dimension.

{\emph{How can practitioners assess the risk given the varying bias magnitudes?}} Towards this end, we derived analytical bounds (Propositions \ref{prop:bias-mean}--\ref{prop:bias-var}) which, while based on simplifying assumptions, are conveniently related to the observable lag-1 autocorrelation $\psi$. Simulations are then used to confirm that these bounds hold robustly (even with GARCH models) -- but these bounds are highly conservative. This is shown in Figure \ref{fig:sim_bias_bounds}. 

Figure \ref{fig:empirical_bounds} shows empirically that the bounds also appear to hold generally, but the relationship is noisy, and violations occur near $\psi \approx 0$; where unmodelled effects and estimation error may dominate the small bias. The discrepancy between empirical and simulation trendlines relating bias-to-bound ratios with the bounds themselves further highlights that the full autocovariance structure, not just the lag-1 autocorrelation, drives the precise bias. In particular, assets may have positive lag-1 autocorrelation but have a negative autocovariance over the window. 
Therefore, the bound's primary utility is as a cautious heuristic: a small bound (low $|\psi|$) suggests low potential bias, making IID resampling less risky; while a large bound signals danger, even if the actual bias might be smaller due to the bound's conservativeness.

For situations where bias is a concern, block resampling offers a direct mitigation strategy. Some temporal structure is preserved by resampling blocks of consecutive observations instead of individual points. Figure \ref{fig:sim_bias_by_blocksize} indicates that bias decreases sufficiently rapidly as block size increases from $b=1$ (IID), suggesting the effectiveness of even minimal structure preservation. However, the reduction in differences on empirical data through increasing block size was much slower. Furthermore, block resampling introduces a trade-off, as larger blocks reduce the effective number of independent samples, potentially limiting the variance reduction achievable through bootstrap aggregation.

In summary, our work reveals that while IID resampling bias is intricately linked to the full return autocovariance structure over the estimation window, its magnitude can range from modest (especially for the Sharpe Ratio due to offsetting effects) to potentially significant relative to estimation noise. Observable first-lag autocorrelation provides a useful, though conservative, heuristic for gauging this risk, and methods like block resampling offer clear mitigation pathways by partially restoring the temporal dependencies that IID resampling disrupts.









\section{Conclusion}\label{sec:conclusion}

Accurate performance evaluation via backtesting is fundamental in portfolio selection research, yet the standard approach using walk-forward cross-validation is often unreliable due to limited data. Resampling methods offer a potential means to increase the use of available data, but understanding their properties is critical to their use. This paper investigated the bias introduced when IID resampling (a canonical resampling approach) is applied to backtesting rolling-window mean-variance portfolios (a canonical benchmark portfolio rule) on temporally dependent financial data.

Our first main contribution is the explanation of the bias mechanism. We demonstrate analytically and through simulation that IID resampling introduces bias primarily by disrupting the first-order autocovariance structure of asset returns. This disruption largely affects estimates of the mean and variance of portfolio rule returns by breaking the train-test dependence structure. Notably, the final Sharpe Ratio bias often benefits from partially offsetting these component biases, explaining why the impact of second-order dependence (e.g., from GARCH) on SR bias was often muted in our findings.

Our second main contribution is the derivation of analytical bounds for this bias and a resulting practical heuristic. The bounds confirm that the bias magnitude is strongly linked to first-lag return autocorrelation, which is observable. This leads to the heuristic: historical first-lag autocorrelation is a crucial indicator of the potential severity of bias. Our simulations and empirical analysis indicate this bias can be non-negligible relative to typical backtest estimation noise, particularly for high-dimensional portfolios and when persistence in returns is high.

Therefore, we recommend caution when using IID resampling for backtesting inference tasks where bias is critical. Its applicability appears limited to scenarios where asset return persistence is low. However, IID resampling may be justified by a substantial reduction in variance for tasks where bias is not critical. We believe future work should first investigate the variance of IID resampled backtests. Other avenues of research are investigations into resampling methods that preserve temporal structure and extensions to broader classes of portfolio rules. 

\section*{Acknowledgements}
Thanks to ...

\section*{Generative AI Disclosure}
ChatGPT o4-mini, and Gemini 2.5 were used in the preparation of this manuscript for developing code templates, and editing written drafts. 

\bibliographystyle{elsarticle-harv} 
\bibliography{resampling}

\clearpage
\newpage

\appendix

\setcounter{page}{1}    
\pagenumbering{arabic}

\section{Variance Reduction through Aggregation}
A Bagged Sharpe Ratio estimate is defined as the average of $K$ bootstrapped Sharpe Ratio estimates:
$$
\hat{\theta}^{B} = \tfrac{1}{K}\sum_{k=1}^K \hat{\theta}_k^* 
$$
where bootstrapped Sharpe Ratio estimates are generated through IID resampling of asset returns. 
The average of many estimators has lower variance than a single estimator (``diversification"):
\begin{align}
    \operatorname{var}\left(\hat{\theta}^{B} \right) = \operatorname{var}\left(\tfrac{1}{K}\sum_{k=1}^K\hat{\theta}_k^*\right) = \tfrac{1}{K} \operatorname{var}\left(\hat{\theta}\right) + \tfrac{1}{K^2}\sum_{\substack{j=1 \\ k \neq j}}^K\operatorname{cov}\left(\hat{\theta}_j^*, \hat{\theta}_k^* \right).  \nonumber
\end{align}
If $\operatorname{cov}\left(\hat{\theta}_j^*, \hat{\theta}_k^* \right) = \rho \operatorname{var}\left(\hat{\theta}_h^*\right)$ is constant for all $i,j$, we have
$$
\operatorname{var}\left(\hat{\theta}^{B} \right) = \tfrac{1}{K}\operatorname{var}\left(\hat{\theta}\right) + \tfrac{K-1}{K}\rho\operatorname{var}\left(\hat{\theta}\right)
$$

Then, as we increase the number of bootstrap samples,  $$\lim_{K \rightarrow \infty} \operatorname{var}\left(\hat{\theta}^{B} \right) = \rho\operatorname{var}\left(\hat{\theta}\right).$$
The limiting variance of the bootstrap aggregated Sharpe ratio is $\rho \times 100 \%$ of the standard backtest Sharpe ratio variance.

\section{Bias Expressions}
We prove the expressions for $\text{bias}\left(\hat{\mu}_{p}^{*}\right)$ and $\text{bias}\left(\hat{\sigma}_p^{*2}\right)$ under the assumptions in Section \ref{ssec:assumptions}.

Firstly, for $\text{bias}\left(\hat{\mu}_{p}^{*}\right)$, we have:
$$
\begin{aligned}
\text{bias}\left(\hat{\mu}^{*}_p\right) &= \E\left[R^{*}_{p,t+1}\right] - \E\left[R_{p,t+1}\right] \\
&= \E[w^{*}_t]\E[R^{*}_{t+1}] - \left({\E[w_t]\E[R_{t+1}] + \text{cov}(w_t,R_{t+1})}\right) \\
&= -\text{cov}(w_t,R_{t+1})
\end{aligned}
$$
When the variance is known, the expectations $\E[w_t]$ and $\E[R_{t+1}]$ are the same whether data is iid or temporally dependent, {\it i.e.} $\E[w_t] = \E[w_t^{*}]$ and $\E[R_{t}] = \E[R_t^*]$. This is because the autocorrelation does not affect the mean of the risky asset, and then we have that $E \left[ R^{*}_{t+1} \right] = E \left[ R_{t+1} \right]$, so that $E\left[\hat{\mu}^{*}_t\right] = E\left[\hat{\mu}_t\right]$, and finally $E[w_t] = E[w_t^{*}]$. This is not true when the variance is unknown. 

Next, for $\text{bias}\left(\hat{\sigma}_p^{*2}\right)$, we have:
\begin{align}
\text{bias}\left(\hat{\sigma}_p^{*2}\right) &= \sigma_{p}^{*2}-\sigma_{p}^{2}  \nonumber \\
&=\text{var}\left( R_{p}^* \right)- \text{var}\left( R_{p} \right) \nonumber  \\
&= \E[R_{p}^{*2}]- \mu_{p}^{*2}-\E[R_{p}^{2}]+\mu_{p}^{2} \nonumber\\
&= \E \left[ w^{*2}_t \right]\E[R^{*2}_{t+1}] - \E \left[ w^{*}_t \right]^{2}\E\left[R^{*}_{t+1}  \right]^{2}\nonumber \\
&\qquad - \left(   \E \left[ w_t^{2} \right]\E[R_{t+1}^{2}] + \operatorname{cov}\left(w_{t}^{2},R_{t+1}^{2}\right)\right) \nonumber \\
&\qquad +   \E \left[ w_t \right]^{2}\E[R_{t+1}]^{2} + 2\E \left[ w_t \right]\E\left[ R_{t+1} \right] \operatorname{cov}\left(w_{t},R_{t+1}\right) \nonumber\\
&\qquad \qquad +\operatorname{cov}^{2}\left(w_{t},R_{t+1}\right). \nonumber
\end{align}
We have (by adding and subtracting $\E \left[ w^{*}_t \right]^{2}\E\left[R^{*2}_{t+1}  \right]$),
\begin{equation}\label{eqn:varw}
\begin{aligned}
&\E \left[ w^{*2}_t \right]\E[R^{*2}_{t+1}] - \E \left[ w^{*}_t \right]^{2}\E\left[R^{*}_{t+1}  \right]^{2}  \\
&= \E \left[ w^{*}_t \right]^{2}\text{var}\left( R_{t+1} \right) + \E \left[ R_{t+1}^{2} \right]\text{var}\left(w^{*}_{t}  \right).  
\end{aligned} 
\end{equation}
Similarly  (by adding and subtracting $\E \left[ w_t \right]^{2}\E\left[R^{2}_{t+1}  \right]$), we have
\begin{equation}\label{eqn:varwstar}
\begin{aligned}
& \E \left[ w^{2}_t \right]\E[R^{2}_{t+1}] - E \left[ w_t \right]^{2}\E\left[R_{t+1}  \right]^{2} \\
&= \E \left[ w_t \right]^{2}\text{var}\left( R_{t+1} \right) + \E \left[ R_{t+1}^{2} \right]\text{var}\left(w_{t}  \right). 
\end{aligned}    
\end{equation}
Therefore, \ref{eqn:varw} - \ref{eqn:varwstar} is $\E \left[ R_{t+1}^{2} \right]\left({ \text{var}\left( w^{*}_{t} \right) - \text{var}(w_{t})}\right)$ because when the variance is known then $E \left[ w_t \right]^{2} - E \left[ w^{*}_t \right]^{2}=0$. Now, 
$$
\begin{aligned}
&2\E \left[ w_t \right]\E\left[ R_{t+1} \right] \operatorname{cov}\left(w_{t},R_{t+1}\right)+\operatorname{cov}^{2}\left(w_{t},R_{t+1}\right) \\
&= \operatorname{cov}\left(w_{t},R_{t+1}\right) \left(\operatorname{cov}\left(w_{t},R_{t+1} \right) + 2 \E[w_{t}]\E \left[ R_{t+1} \right]  \right)  \\
&=\tfrac{1}{2}\operatorname{cov}\left(w^2_{t},R_{t+1}^{2}\right)
\end{aligned}
$$
where equality between the second and third lines is true for Gaussian returns with known variance. 

Putting it all together, 
$$
\text{bias}\left(\hat{\sigma}_p^{*2}\right) = \E \left[ R_{t+1}^{2} \right]\left({\text{var}(w^{*}) - \text{var}(w)}\right) - \tfrac{1}{2}\operatorname{cov}\left(w^{2},R_{t+1}^{2}\right).
$$

\section{Preliminaries}
For the derivations below, we define the following sums:
\begin{align}
A_{n,\phi} &= \tfrac{1}{n}\sum_{i=t-n+1}^t\phi^{|t+1-i|} =\tfrac{1}{n} \sum_{k=1}^n\phi^k,  \label{eqn:appC.A}\\
B_{n,\phi} &= \tfrac{1}{n^2}\sum_{i=1}^n\sum_{\substack{ j \neq i}}^n\phi^{|i-j|}, \\
\end{align}

The sums $A_{n,\phi}$, $B_{n,\phi}$ are bounded for $\phi \in [0,1)$ and all $n$. We use the following bounds in our derivations for the bias in the mean and variance.

Since $\phi^{k}\leq\phi$ for all $k \geq 1, \phi \in [0,1)$, we have 
$$
\begin{aligned}
A_{n,\phi}&: \quad 0\leq   \tfrac{1}{n}\sum_{k=1}^n\phi^k \lt \tfrac{1}{n}n\phi \lt \phi, \\
B_{n,\phi}&: \quad 0 \leq \tfrac{1}{n^2}\sum_{i=1}^n\sum^n_{\substack{j=1, j\neq i}}\phi^{|i-j|} \lt \phi. \\
\end{aligned}
$$

Our derivations use the following useful property of multivariate Gaussian variables. If $X$ and $Y$ are jointly normally distributed, with non-zero means, we have:
$$\label{eq:mvprop3}
\text{cov}(X^2,Y^2) = 2\text{cov}(X,Y)\left(\text{cov}(X,Y) +2\E[X]\E[Y]\right)
$$


\section{Bounding the resampled performance}\label{appsec:mean_bound}

Here we consider the bounds on the normalised resampled portfolio performance bias: $\tfrac{1}{\mu_{p}} \text{bias}\left(\hat{\mu}_{p}^{*}\right)$.

\subsection{Resampled performance bias bounds}

We start by considering the bounds on the portfolio performance bias $\text{bias}\left(\hat{\mu}_{p}^{*}\right)$:
\begin{align}
\text{bias}\left(\hat{\mu}^{*}_p\right) &= -\text{cov}(w_t,R_{t+1})
= -\tfrac{1}{\sigma_{R}^{2}}\text{cov}(\hat{\mu}_{t},R_{t+1}). \label{eqn:appD1.bias}
\end{align}
Assuming rolling sample estimators:
\begin{align}
	\hat{\mu}_t = \tfrac{1}{n}\sum_{i=t-n+1}^tR_{i} &= \tfrac{1}{n}\sum_{i=t-n+1}^t\mu_{i} + \epsilon_i = \bar{\mu}_t + \bar{\epsilon}_t. \label{eqn:appD1.mean}
\end{align}
Additionally, we have $R_{t+1} = \mu_{t+1}+\epsilon_{t+1}$. Therefore, we have:
\begin{align}
-\tfrac{1}{\sigma_{R}^{2}}\text{cov}(\hat{\mu}_{t},R_{t+1}) =-\tfrac{1}{\sigma_{R}^{2}}\text{cov}(\bar{\mu}_t, \mu_{t+1}) \label{eqn:appD1.cov}
\end{align}
since the noise terms $\epsilon_{t}$ are IID. Now, focusing on the first term we have:
$\text{cov}(\mu_t, \mu_{t-k}) = \phi^k\sigma^2_\mu$. Therefore using Eqn. \ref{eqn:appC.A} and Eqn. \ref{eqn:appD1.cov}
$$
\text{cov}(\bar{\mu}_t, \mu_{t+1}) = \tfrac{\sigma_\mu^2}{n}\sum_{k=1}^n\phi^k =\sigma_\mu^2A_{n,\phi}.
$$
Which gives an expression for the bias:
\begin{align}
\text{bias}\left(\hat{\mu}_p^{*}\right)=-\frac{1}{\gamma} \frac{\sigma_{\mu}^{2}}{\sigma_{R}^{2}} A_{n,\phi}.
\end{align}
If we assume $\phi>0$, then we can bound the covariance:
$$
0  \leq\text{cov}(\bar{\mu}_t, \mu_{t+1}) \leq\sigma_{\mu}^2\phi
$$
where the upper bound is attained when $n=1$. This allows us to bound the portfolio performance bias. Furthermore, we define $\psi = \text{corr}(R_t, R_{t+1})$. Then, assuming constant covariance
$$
\begin{aligned}
\psi &= \tfrac{1}{\sigma^2_R}\text{cov}(R_t, R_{t+1})\\
     &= \tfrac{1}{\sigma^2_R}\text{cov}(\mu_t + \epsilon_t, \mu_{t+1} + \epsilon_{t+1}) \\
     &= \tfrac{1}{\sigma^2_R}\text{cov}(\mu_t, \mu_{t+1}) = \tfrac{\sigma_{\mu}^2}{\sigma^2_R}\phi.
\end{aligned}
$$

Therefore, 
$$
-\tfrac{1}{\gamma} \psi \leq -\tfrac{1}{\gamma} \tfrac{\sigma_{\mu}^{2}}{\sigma_{R}^{2}} \phi \leq \text{bias}\left(\hat{\mu}_p^{*}\right) \leq 0
$$
where the lower bound is attained when $n=1$.

\subsection{Bounding portfolio performance}

We start by solving for average portfolio performance $\mu_{p}$ under the original DGP.
\begin{align}
\mu_{p} &= \E[R_{p,t+1}] \nonumber  \\
&= \E[w_{t}]\E[R_{t+1}] + \operatorname{cov}\left(w_{t},R_{t+1}\right) \nonumber  \\
&= \tfrac{1}{\gamma}\tfrac{1}{\sigma_{R}^{2}}\E\left[\hat{\mu}_t\right]\E \left[ R_{t+1} \right] + \tfrac{1}{\gamma}\tfrac{1}{\sigma_{R}^{2}}\operatorname{cov}\left(\hat{\mu}_{t},R_{t+1}\right) \nonumber  \\
&= \tfrac{1}{\gamma}\left( \tfrac{\mu^{2}}{\sigma_{R}^{2}}+\tfrac{\sigma_{\mu}^{2}}{\sigma_{R}^{2}} A_{n,\phi}\right) \leq \tfrac{1}{\gamma}\left( \theta^{2}+\psi \right) . 
\end{align}

Similarly, we have:
\begin{align}
\mu^{*}_{p} &= \E[R^{*}_{p,t+1}] \nonumber  \\
&= \E[w^{*}_{t}]\E[R^{*}_{t+1}] + \operatorname{cov}\left(w^{*}_{t},R^{*}_{t+1}\right) \nonumber  \\
&= \tfrac{1}{\gamma}\tfrac{1}{\sigma_{R}^{2}}\E\left[\hat{\mu}^{*}_t\right]\E \left[ R^{*}_{t+1} \right] \nonumber  \\
&= \tfrac{1}{\gamma}\left( \tfrac{\mu^{2}}{\sigma_{R}^{2}}\right) = \tfrac{1}{\gamma} \theta^{2}  . 
\end{align}




\section{Bounding the resampled portfolio variance}\label{appsec:var_bound}

Here we consider the bounds on the resampled portfolio performance variance $\tfrac{1}{\sigma^{2}_{p}}\text{bias}\left(\hat{\sigma}_p^{*2}\right)$.

\subsection{Resampled performance variance bias}

We showed earlier that bias in the variance of the resampled portfolio performance is: 
$$
\text{bias}\left(\hat{\sigma}_p^{*2}\right) = \E \left[ R_{t+1}^{2} \right]\left(\text{var}(w^{*}) - \text{var}(w)\right) - \tfrac{1}{2}\operatorname{cov}\left(w^{2},R_{t+1}^{2}\right).
$$
First, we find a bound for $\text{var}(w^*) - \text{var}(w)$. This follows from:
$$
\begin{aligned}
\text{var}\left( w \right) &= \tfrac{1}{\gamma^{2}\sigma_{R}^{4}}\text{var}\left(\hat{\mu}_{t}  \right)  \\
&= \tfrac{1}{\gamma^{2}\sigma_{R}^{4}} \tfrac{1}{n^2}\text{var}\left(\textstyle\sum_{i=t-n+1}^t R_i\right) \\
&= \tfrac{1}{\gamma^{2}\sigma_{R}^{4}} \left( \tfrac{1}{n}\text{var}(R_i) + \tfrac{1}{n^2}\textstyle\sum_{i, j \neq i} \text{cov}\left(R_i, R_j\right) \right) \\
&= \tfrac{1}{\gamma^{2}\sigma_{R}^{4}} \left( \tfrac{1}{n}\text{var}(R_i) + \tfrac{1}{n^2}\textstyle\sum_{i,j \neq i} \text{cov}\left(\mu_i, \mu_j\right) \right) \\
&= \tfrac{1}{\gamma^{2}\sigma_{R}^{4}} \left(    \tfrac{\sigma^2_R}{n} + \sigma^2_{\mu}B_{n,\phi}\right) \\
& \leq \tfrac{1}{\gamma^{2}\sigma_{R}^{4}} \left( \tfrac{\sigma_{R}^{2}}{n} + \sigma_{\mu}^{2} \phi \right) \\
&\leq  \tfrac{1}{\gamma^{2}\sigma_{R}^{2}} \left( \tfrac{1}{n} + \tfrac{\sigma_{\mu}^{2}}{\sigma_{R}^{2}}\phi  \right).
\end{aligned}
$$
Now, $\text{var}\left( w^{*} \right)$ is the same as $\text{var}(w)$ except that $\operatorname{cov}\left(R_{i},R_{j}\right) = 0$ for $i \neq j$. Therefore, $B_{n,\phi}=0$, and we have
$$
\text{var}(w^*) = \frac{1}{\gamma^{2}} \frac{\sigma_{R}^{2}}{n},
$$
Furthermore, 
$$
\begin{aligned}
\text{var}(w^*) - \text{var}(w) &=-\frac{1}{\gamma^{2}} \sigma_{\mu}^{2} B_{n,\phi} \geq -\frac{1}{\gamma^{2}} \sigma_{\mu}^{2}\phi.
\end{aligned}
$$
Using this result, we can bound bias in the resampled portfolio performance variance $\text{bias}\left(\hat{\sigma}_p^{*2}\right)$:
$$
\begin{aligned}
\text{bias}&\left({\hat{\sigma}_p^{*2}}\right) = \E \left[ R_{t+1}^{2} \right]\left(\text{var}(w^{*}) - \text{var}(w)\right) - \tfrac{1}{2}\operatorname{cov}\left(w^{2},R_{t+1}^{2}\right) \\ 
&=\tfrac{1}{\gamma^{2}}\tfrac{1}{\sigma_{R}^4}\left[\E \left[ R_{t+1}^{2}   \right] \left(\text{var}\left( \hat{\mu}_{t}^{*} \right) - \text{var}\left( \hat{\mu}_{t} \right)\right) - \tfrac{1}{2}\operatorname{cov}\left(\hat{\mu}_{t}^{2}, R_{t+1}^{2}\right)\right] \\
&= \tfrac{1}{\gamma^{2}}\tfrac{1}{\sigma_{R}^4} \left[{\E \left[ R_{t+1}^{2} \right]\left(\text{var}\left( \hat{\mu}_{t}^{*} \right) - \text{var}\left( \hat{\mu}_{t} \right)\right) } \right. \\ 
 &~~~~~~~\left.{-\operatorname{cov}\left(\hat{\mu}_{t},R_{t}\right)\left( \operatorname{cov}\left(\hat{\mu}_{t},R_{t}\right) + 2 \E\left[\hat{\mu}_t\right] \E[R_{t+1}]\right) }   \right] \\
&= \tfrac{1}{\gamma^{2}}\tfrac{1}{\sigma_{R}^4} \left[{ -\left( \mu^{2}+\sigma_{R}^{2} \right)\sigma^2_{\mu}B_{n,\phi}  -\sigma_\mu^2 A_{n,\phi}\left( \sigma_\mu^2 A_{n,\phi} +2\mu^{2} \right) }    \right] \\
&\geq \tfrac{1}{\gamma^{2}}\tfrac{1}{\sigma_{R}^4}\left[{ -\left( \mu^{2}+\sigma_{R}^{2} \right)\sigma^2_{\mu}\phi - \sigma_{\mu}^{2}\phi\left( \sigma_{\mu}^{2}\phi+2\mu^{2} \right)}  \right] \\
&\geq -\tfrac{1}{\gamma^{2}} \tfrac{\sigma_{\mu}^{2}}{\sigma_{R}^{2}} \phi\left( \tfrac{3\mu^{2}}{\sigma_{R}^{2}}+\tfrac{\sigma_{\mu}^{2}}{\sigma_{R}^{2}}\phi + \tfrac{\sigma_{R}^{2}}{\sigma_{R}^{2}} \right) \\
&\geq -\tfrac{1}{\gamma^{2}}\psi\left( 3\theta^{2} + \psi+1 \right)  \\
&\geq -\tfrac{C}{\gamma^{2}}\psi  \quad \text{where} \quad C= \psi\left( 3\theta^{2} + \psi+1 \right).
\end{aligned}
$$

\subsection{Bounding the portfolio performance variance}

Furthermore, we have:
$$
\begin{aligned}
\sigma^{*2}_p &= E[R^{*2}_p]-\mu^{*2}_p\\
&= E[w^{*2}_t]E[R^2_{t+1}]-\mu^{*2}_p\\
&\approx \frac{1}{\gamma^2}\frac{\mu^2}{\sigma^4_R}\left(\sigma_{R}^2+\mu^2\right)-\mu^{*2}_p \\
&= \frac{1}{\gamma^2}\frac{\mu^2}{\sigma^2_{R}} + \frac{1}{\gamma^2}\frac{\mu^4}{\sigma^4_{R}} -\left(\frac{1}{\gamma}\frac{\mu^2}{\sigma_{R}^2}\right)^2 \\
&= \frac{1}{\gamma^2}\frac{\mu^2}{\sigma^2_{R}} = \frac{1}{\gamma^2}\theta^2
\end{aligned}
$$






\section{Bounding the Resampled Sharpe Ratio}

\subsection{Analytical Bound}\label{appssec:analytical_bound}
Substituting Propositions \ref{prop:bias-mean} and \ref{prop:bias-var} into the bias for Sharpe Ratio equation \ref{eq:bias_sharpe}:
\begin{equation}
\begin{split}
\text{bias}\left(\widehat{\Theta}^{*}_p\right) =\frac{1}{\theta}\Bigl(&- \tfrac{\sigma_\mu^2}{\sigma_R^2}A_{n,\phi}
+ \tfrac12\bigl((\theta^2+1)\tfrac{\sigma_\mu^2}{\sigma_R^2}B_{n,\phi}\\
&\quad+ \tfrac{\sigma_\mu^2}{\sigma_R^2}A_{n,\phi}
\bigl(\tfrac{\sigma_\mu^2}{\sigma_R^2}A_{n,\phi}+2\theta^2\bigr)\bigr)\Bigr)\\
&= \tfrac{\sigma_\mu^2}{2\theta\,\sigma_R^2}
\Bigl(\tfrac{\sigma_\mu^2}{\sigma_R^2}A_{n,\phi}^2
+2(\theta^2-1)A_{n,\phi}
+(\theta^2+1)B_{n,\phi}\Bigr).
\end{split}
\end{equation}

Substituting the Bounds $A_{n,\phi} \leq \phi$ and $B_{n,\phi} \leq \phi$ gives the bound:
$$
\text{bias}\left(\widehat{\Theta}^{*}_p\right) \leq \frac{1}{2}\frac{\psi}{\theta}(C-2)
$$

\subsection{Numerical Bound}\label{appssec:num_bound}

We investigate the behaviour of the true Sharpe Ratio with the window size $n$. The true Sharpe Ratio $\Theta_p$ is a non-linear and can be non-monotonic in $n$. This is because the true portfolio rule variance $\sigma_p^2$ is non-linear and non-monotonic in $n$. In particular, $\sigma_p^2$ can increase initially for small values of $n$, typically $n\leq5$, and then decreases monotonically afterwards. 

This non-monotonic behaviour stems from the sum $B_{n,\phi}$, which is defined as a double sum scaled by $1/n^2$. If we could replace the $n^2$ by $1/n(n-1)$, the variance would decrease monotonically. Since $n(n-1)<n^2$, it would also amplify its dividend, making the variance larger. We use this in our bounding arguments to simplify the algebra, which is why our bound is loose as seen below. If the variance is monotonically decreasing, we also have that $\Theta_{n,p}$ is monotonically decreasing, and $\Theta_{1,p}$ would be the maximum and given by the numeric bounds.

However, the variance is not monotonically decreases, and the variance bound is loose. Overstating the maximum variance causes the $\Theta_{n,p}$ to be larger than the numeric bound. However, through large experimentation, we found that the bound was valid for $n>10$. While this is not rigorous, we believe it is a useful conjecture that the bound is valid $n>10$.

\begin{figure*}[h!]
    \centering
    \includegraphics[width=0.8\textwidth]{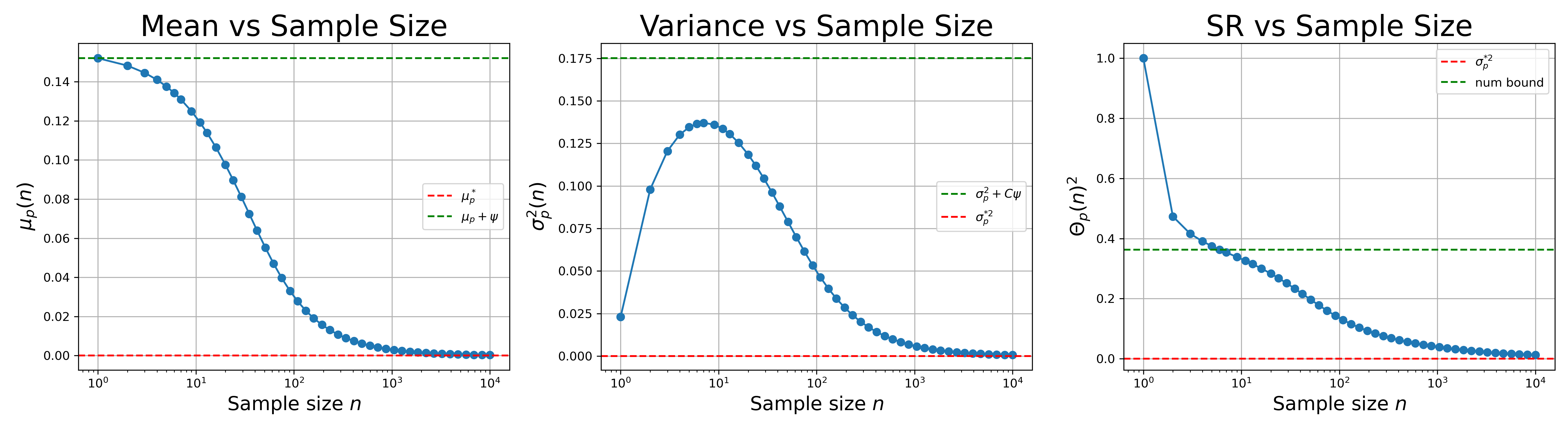}
    \caption{The portfolio rule's true mean, variance, and Sharpe Ratio $\mu_p, \sigma^2_p, \Theta_p$ as a function of window size $n$. The dashed lines for the mean and variance are the derived bounds. The dashed line for the Sharpe Ratio is the numerically derived bound. As can be seen, the Sharpe Ratio is below the bound for practical window sizes $n > 10$. We repeat this examination many times with different parameters, with the same behaviour.  }
    \label{fig:sim_analytical_vs_numeric_by_n}
\end{figure*}

\subsection{Empirical Comparison}\label{appsec:emp_comparison}

We compare the analytical and numerical bound for the MonteCarlo and Historical simulation.

\begin{figure*}[h!]
    \centering
    \includegraphics[width=0.7\textwidth]{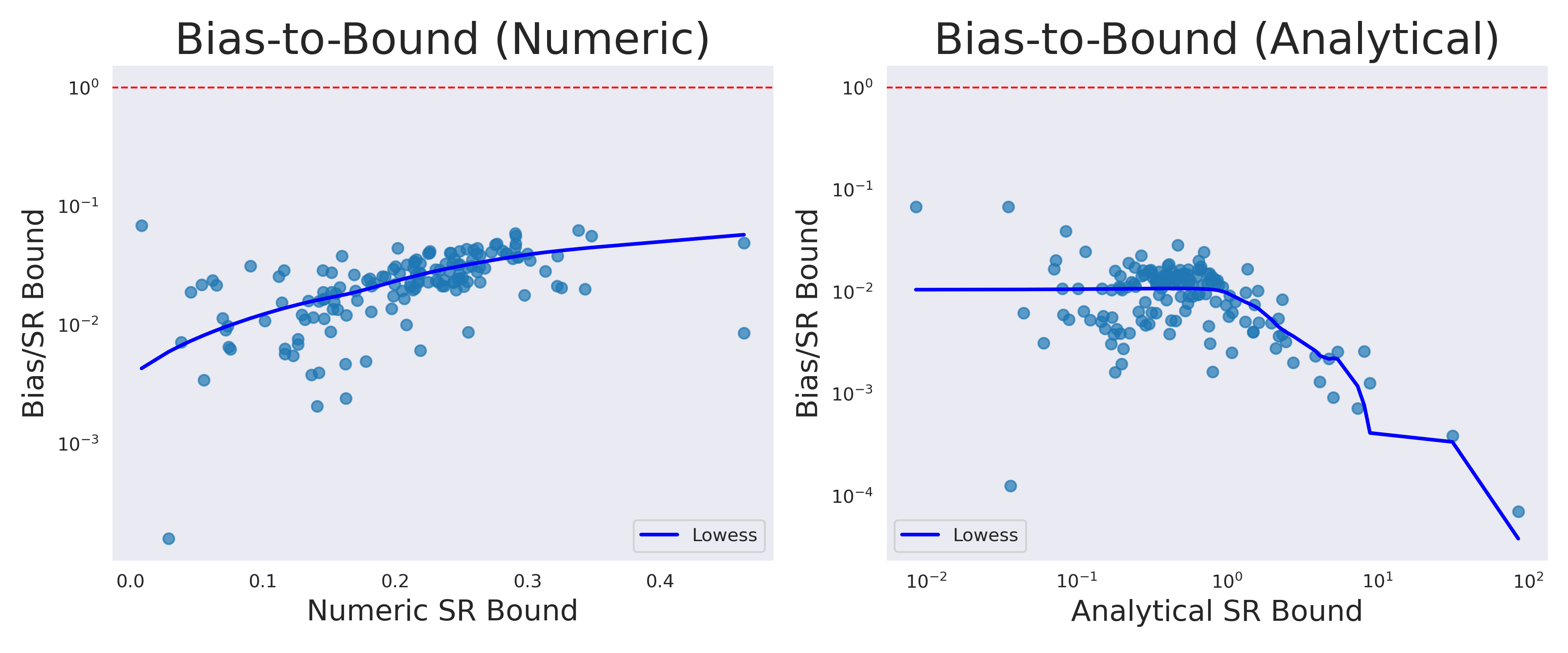}
    \caption{Comparison of numerical and analytically derived bounds for simulated data. The x-axis of the analytical bounds is also on a log-scale. For small values of $\theta$, the bound explodes with values reaching close to $10^3$. }
    \label{fig:sim_analytical_vs_numeric_bounds}
\end{figure*}

\begin{figure*}[h!]
    \centering
    \includegraphics[width=0.7\textwidth]{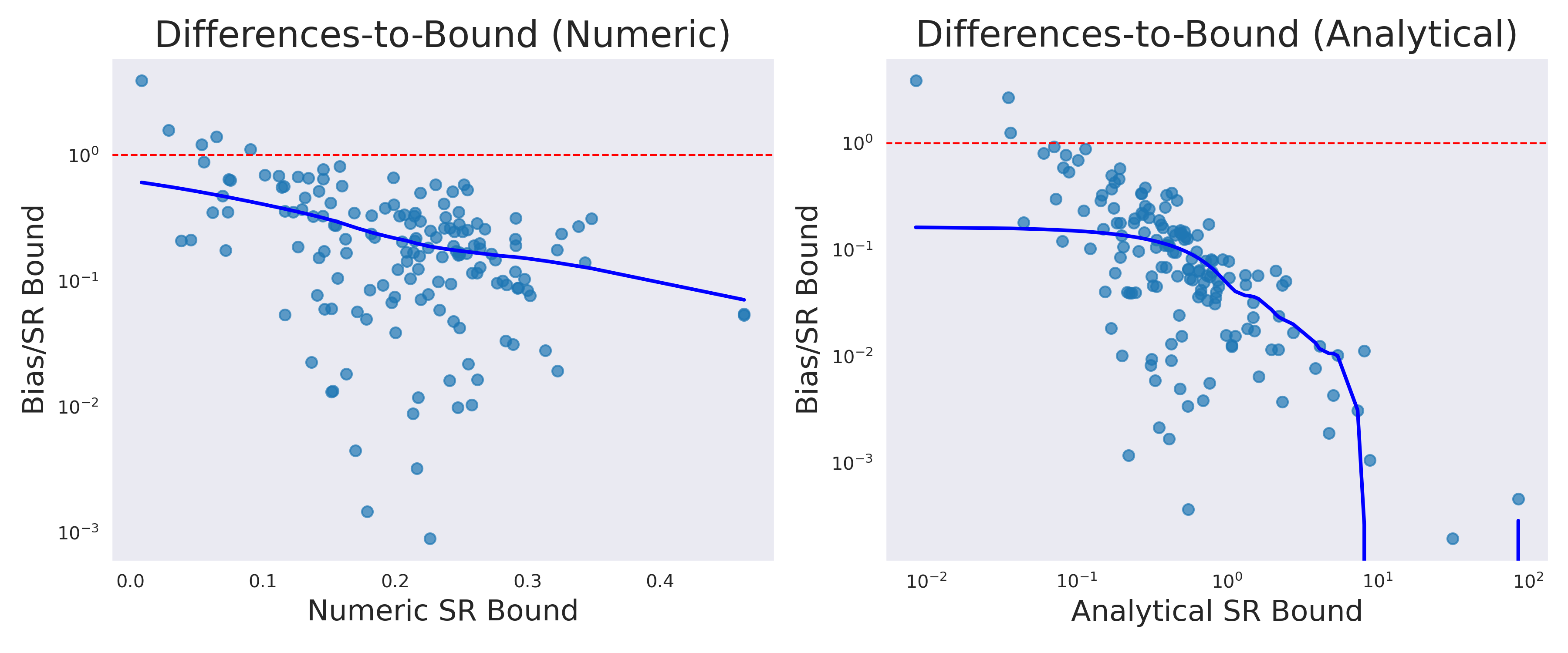}
    \caption{Comparison of numerical and analytically derived bounds for historical data. The x-axis of the analytical bounds is also on a log-scale. For small values of $\theta$, the bound explodes with values reaching close to $10^3$. }
    \label{fig:emp_analytical_vs_numeric_bounds}
\end{figure*}

\section{The behaviour of the average autocorrelation sums}\label{appsec:assymptotic_sums}

We briefly investigate the behaviour of the sums $A_{n,\phi}$ and $B_{n,\phi}$ and their relationship, which is important in understanding the offsetting mechanism. 

As the window size $n\rightarrow\infty$ increases, both sums tend to $0$:
$$
\begin{aligned}
    \lim_{n \to \infty}A_{n,\phi} = \left(\lim_{n \to \infty} \tfrac{1}{n} \right)\left(\tfrac{1}{1-\phi} \right) = 0
\end{aligned}
$$

We analyse the finite sample and asymptotic behaviour of the ratio $R_{n,\phi} = B_{n,\phi}/A_{n,\phi}$ for $0 < \phi < 1$, 

First, we simplify $A_{n,\phi}$. The sum is a standard finite geometric series:
\begin{equation}
\sum_{k=1}^n \phi^k = \phi \frac{1-\phi^n}{1-\phi}.
\end{equation}
Therefore,
\begin{equation}
A_{n,\phi} = \frac{1}{n} \frac{\phi(1-\phi^n)}{1-\phi}. \label{eqn:A_final}
\end{equation}

Next, we analyze $B_{n,\phi}$. The sum involved is the sum of the off-diagonal elements of an $n \times n$ matrix $M$ with entries $M_{ij} = \phi^{|i-j|}$. Let this sum be $O_n$:
\begin{equation}
O_n = \sum_{i=1}^n\sum^n_{\substack{j=1 \\ j\neq i}}\phi^{|i-j|}.
\end{equation}
We can find $O_n$ by considering the total sum $S_n = \sum_{i=1}^n\sum_{j=1}^n \phi^{|i-j|}$ and subtracting the sum of the diagonal elements $D_n = \sum_{i=1}^n \phi^{|i-i|} = \sum_{i=1}^n 1 = n$. Thus, $O_n = S_n - n$.

We calculate $S_n$ by summing the row sums $R_i = \sum_{j=1}^n \phi^{|i-j|}$.
\begin{align*}
R_i &= \sum_{j=1}^{i-1} \phi^{i-j} + \phi^0 + \sum_{j=i+1}^{n} \phi^{j-i} \\
    &= \sum_{k=1}^{i-1} \phi^k + 1 + \sum_{k=1}^{n-i} \phi^k \\
    &= 1 + \frac{\phi(1-\phi^{i-1})}{1-\phi} + \frac{\phi(1-\phi^{n-i})}{1-\phi}.
\end{align*}
Summing $R_i$ over $i=1,\dots,n$:
\begin{align*}
S_n &= \sum_{i=1}^n \left( 1 + \frac{\phi}{1-\phi} [ (1-\phi^{i-1}) + (1-\phi^{n-i}) ] \right) \\
    &= n + \frac{\phi}{1-\phi} \sum_{i=1}^n (2 - \phi^{i-1} - \phi^{n-i}) \\
    &= n + \frac{2n\phi}{1-\phi} - \frac{\phi}{1-\phi} \sum_{i=1}^n (\phi^{i-1} + \phi^{n-i}).
\end{align*}
The two sums in the last term are identical geometric series:
\begin{equation*}
\sum_{i=1}^n \phi^{i-1} = \sum_{k=0}^{n-1} \phi^k = \frac{1-\phi^n}{1-\phi}.
\end{equation*}
Letting $j=n-i$, as $i$ goes from $1$ to $n$, $j$ goes from $n-1$ down to $0$:
\begin{equation*}
\sum_{i=1}^n \phi^{n-i} = \sum_{j=0}^{n-1} \phi^j = \frac{1-\phi^n}{1-\phi}.
\end{equation*}
Substituting these back into $S_n$:
\begin{align*}
S_n &= n + \frac{2n\phi}{1-\phi} - \frac{\phi}{1-\phi} \left( 2 \frac{1-\phi^n}{1-\phi} \right) \\
    &= \frac{n(1-\phi)^2 + 2n\phi(1-\phi) - 2\phi(1-\phi^n)}{(1-\phi)^2}.
\end{align*}
Simplifying the numerator:
\begin{align*}
& n(1-2\phi+\phi^2) + (2n\phi - 2n\phi^2) - (2\phi - 2\phi^{n+1}) \\
= & n - 2n\phi + n\phi^2 + 2n\phi - 2n\phi^2 - 2\phi + 2\phi^{n+1} \\
= & n - n\phi^2 - 2\phi + 2\phi^{n+1} = n(1-\phi^2) - 2\phi(1-\phi^n).
\end{align*}
Thus, the total sum is:
\begin{equation}
S_n = \frac{n(1-\phi^2) - 2\phi(1-\phi^n)}{(1-\phi)^2}. \label{eqn:S_final}
\end{equation}
The sum of off-diagonal elements is $O_n = S_n - n$:
\begin{align}
O_n &= \frac{n(1-\phi^2) - 2\phi(1-\phi^n)}{(1-\phi)^2} - \frac{n(1-\phi)^2}{(1-\phi)^2} \nonumber \\
    &= \frac{n - n\phi^2 - 2\phi + 2\phi^{n+1} - n(1-2\phi+\phi^2)}{(1-\phi)^2} \nonumber \\
    &= \frac{n - n\phi^2 - 2\phi + 2\phi^{n+1} - n + 2n\phi - n\phi^2}{(1-\phi)^2} \nonumber \\
    &= \frac{2n\phi - 2n\phi^2 - 2\phi + 2\phi^{n+1}}{(1-\phi)^2} \nonumber \\
    &= \frac{2n\phi(1-\phi) - 2\phi(1-\phi^n)}{(1-\phi)^2}. \label{eqn:O_final}
\end{align}

Then $B_{n,\phi}$ is given by:
\begin{align*}
B_{n,\phi} &= \frac{1}{n^2} O_n = \frac{1}{n^2} \frac{2n\phi(1-\phi) - 2\phi(1-\phi^n)}{(1-\phi)^2} \nonumber \\
           &= \frac{2\phi(1-\phi)}{n(1-\phi)^2} - \frac{2\phi(1-\phi^n)}{n^2(1-\phi)^2} \nonumber \\
           &= \frac{2\phi}{n(1-\phi)} - \frac{2\phi(1-\phi^n)}{n^2(1-\phi)^2}. \label{eqn:B_final}
\end{align*}
Finally, the ratio $R_{n,\phi} = B_{n,\phi}/A_{n,\phi}$ is:
\begin{align*}
R_{n,\phi} &= \frac{\frac{2\phi}{n(1-\phi)} - \frac{2\phi(1-\phi^n)}{n^2(1-\phi)^2}}{\frac{\phi(1-\phi^n)}{n(1-\phi)}} \\
  &= \left( \frac{2\phi}{n(1-\phi)} - \frac{2\phi(1-\phi^n)}{n^2(1-\phi)^2} \right) \frac{n(1-\phi)}{\phi(1-\phi^n)} \\
  &= \frac{2}{1-\phi^n} - \frac{2(1-\phi^n)}{n(1-\phi)} \frac{1}{(1-\phi^n)} \\
  &= \frac{2}{1-\phi^n} - \frac{2}{n(1-\phi)}.
\end{align*}
This is the exact result for finite $n$.

For the asymptotic behavior as $n \to \infty$, since $0 < \phi < 1$, we have $\lim_{n \to \infty} \phi^n = 0$.
\begin{align*}
\lim_{n \to \infty} R_{n,\phi} &= \lim_{n \to \infty} \left( \frac{2}{1-\phi^n} - \frac{2}{n(1-\phi)} \right) \\
 &= \frac{2}{1 - 0} - 0 = 2.
\end{align*}

\section{Simulation Details}

\subsection{Parameter Specification}\label{appssec:parm_spec}

\begin{figure*}[h!]
    \centering
    \includegraphics[width=0.7\textwidth]{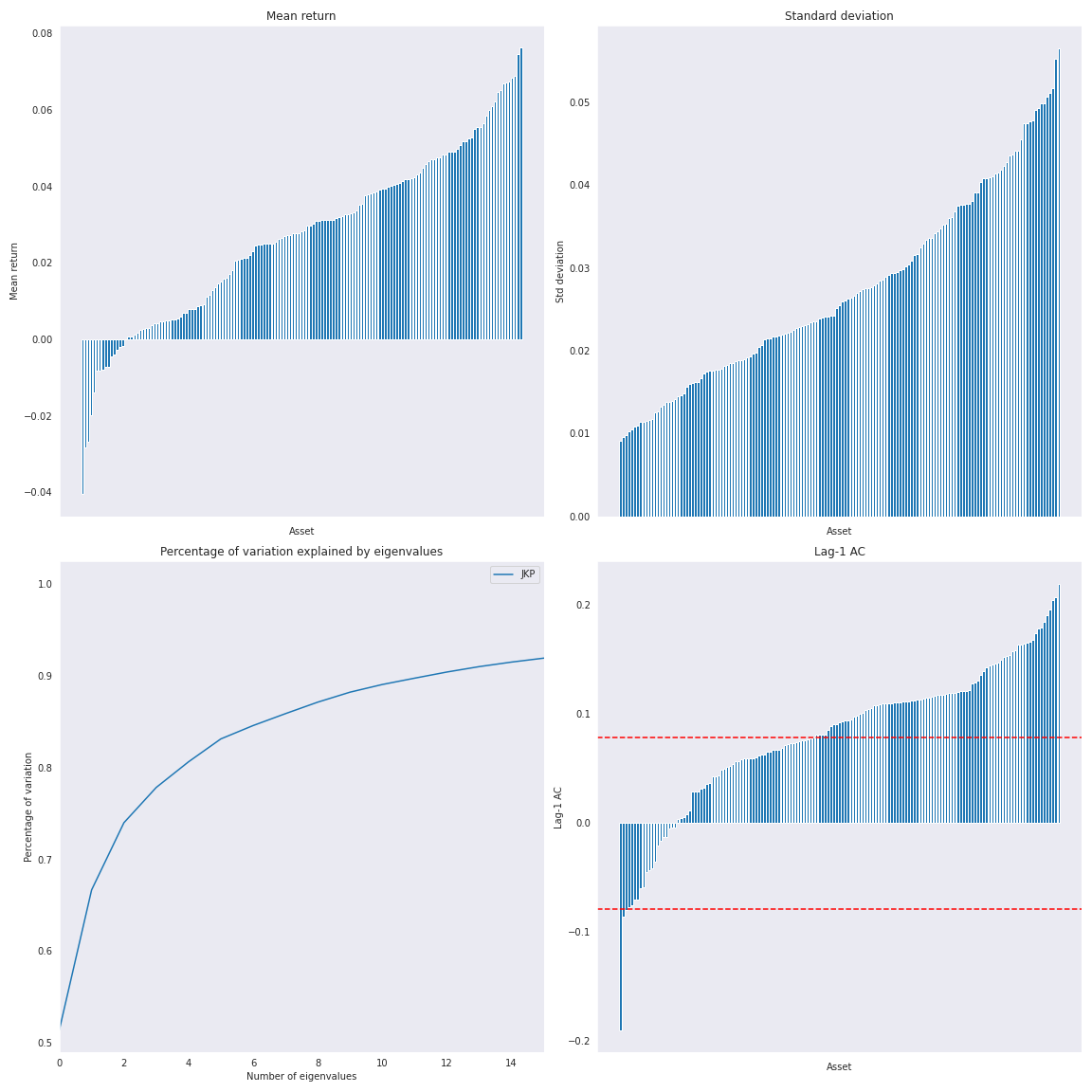}
    \caption{Exploratory Analysis of JKP Data. Mean (top left), Std Deviation (top right), Eigenvalue Structure (bottom left), and Lag-1 Return Auto-correlation of each asset (bottom right). The red horizontal lines on the Autocorrelation plot are thresholds for significance. The mean, variance, and lag-1 autocorrelation statistics are used to specify $\mu, \sigma, \psi$ for each asset in the cross-section.  }
    \label{fig:jkp_exploration}
\end{figure*}

Estimating the time-varying risk premium process is non-trivial, requiring sophisticated estimation procedures. While exact estimation is preferred, it is not required to achieve the main aims of our simulation study: namely, to explore deviations from assumptions of our analytical model. However, we would like magnitudes with some context within the real world. Therefore, we forego a complicated estimation process, but we use an exploratory data analysis to select parameters that are consistent with ranges reported empirically. We use method of moment estimation to specify the unconditional means $\mu_t$, variances $\sigma_t^2$, and autocorrelations of the assets $\psi$. We plot summary statistics (and hence the observable parameters in our simulations) for the JKP data in Figure \ref{fig:jkp_exploration}. 

The unobserved parameters $\phi_t, \sigma_\mu^2, \sigma_\nu^2$ are specified in an ad-hoc manner, which we justify as follows. Firstly, we note that the ratio $\tfrac{\sigma_\mu^2}{\sigma^2_{\mu}}$ is the proportion of the total return variance that is explained by the variation in the risk premium. This is closely related to an $R^2$ value of a regression model. Prediction models are well known to have low $R^2$s out-of-sample, but this is more of a reflection of the prediction model rather than the inherent variation in the risk premium. So, we consider in-sample $R^2$ instead. While this has the converse problem where a high $R^2$ typically reflects overfitting rather than variation in the risk premium, setting $\sigma_{\mu}^2$ to correspond to a high $R^2$ implies that our data has large signal content -- conditions where we expect IID resampling to fail. 

\begin{figure*}[h!]
    \centering
    \includegraphics[width=0.9\columnwidth]{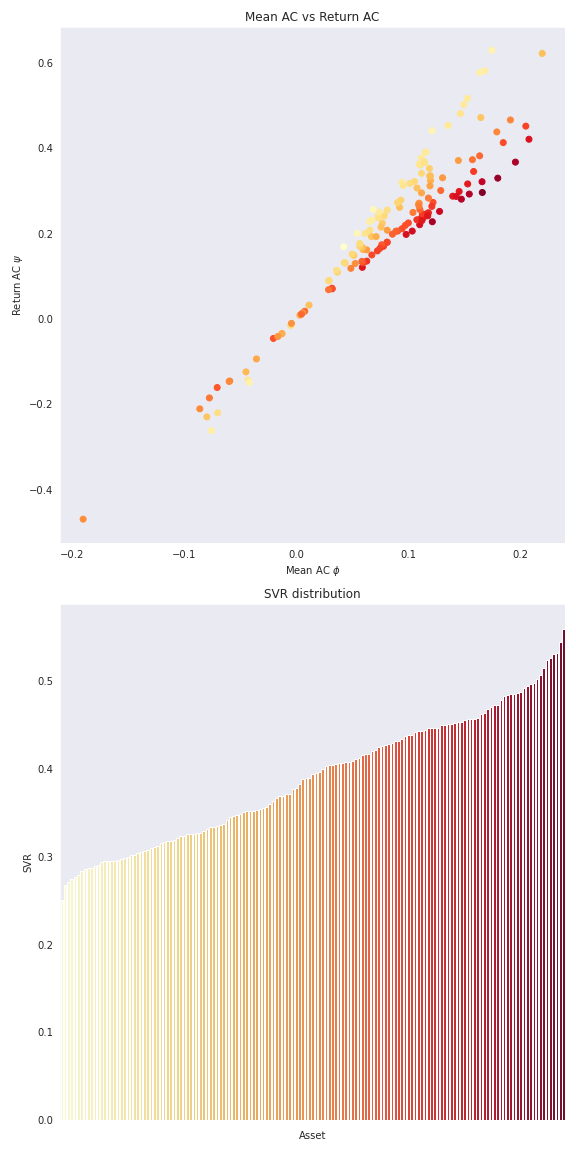}
    \caption{Unobserved Parameters $\phi$ vs $\psi$ (top) and $\text{SVR}=\tfrac{\sigma^2_\mu}{\sigma_R^2}$ (below). $\phi$ and $\psi$ are closely linearly related. The SVR is generated from in-sample $R^2$ values from a VAR(1) model fitted to the data. }
    \label{fig:jkp_phi_svr_parms}
\end{figure*}

Thus, we specify a range of $\sigma_\mu^2$ that includes low signal and high-signal conditions to explore the effect on the bias of the IID resampled backtests. We fit a VAR(1) model to the JKP returns, and use the in-sample $R^2$ together with the $\sigma_R^2$ parameters to specify the range of $\sigma_{\mu}^2$:
$$
    \sigma^2_{\mu} = R^2\sigma^2_R.
$$
We use the relation from Equation \ref{eq:psi} to specify $\phi$:
$$
\psi \tfrac{\sigma_R^2}{\sigma_\mu^2} = \phi.
$$
Lastly, we specify $\sigma_{\nu}^2$ using Equation \ref{eq:sigma_mu}:
$$
\sigma_\nu^2 =\sigma^2_{\mu}(1-\phi^2) 
$$
We plot the distributions of the unobservable parameters in Figure \ref{fig:jkp_phi_svr_parms}.

Our first extension is to examine the effect of GARCH errors. The $\alpha$ and $\beta$ GARCH parameters are specified by fitting univariate AR(1)-GARCH(1,1) models to each asset in the JKP data. The $\omega$ parameter is specified so that the unconditional variance of the return process is the same in the GARCH and constant variance cases. To do this, we specify $\omega$ using the formula for GARCH variance:
$$
\omega = \sigma_\epsilon^2\left(1 - \alpha - \beta\right)
$$
where $\sigma_\epsilon^2 = \sigma_R^2 - \sigma_mu^2$. The sums of the dependence parameters are specified to have a range $\alpha+\beta \in [0.74, 0.98]$. Where the estimated parameters are outside this range, the $\alpha$ value is modified so that the sum is at the range's boundary. 

Secondly, we examine the impact of increasing dimension. Thus far, our simulation DGP treats each asset in isolation as a univariate process. A multivariate specfication requires that we specify covariance matrices $\Sigma_\mu, \Sigma_R, \Sigma_{\nu}$ and autocorrelation matrices $\Psi, \Phi$. For simplicity and consistency with the parameters used in the univariate results, we extend to the multivariate case by assuming diagonal matrices. We faced difficulty with choosing parameters such that all covariance matrices are valid (positive definite) but still consistent with the univariate specification. Thus, we examine covariance matrices with an explicit factor structure separately with parameters that are not consistent with those used in the univariate results we present in Section \ref{ssec:results}. Qualitatively, the behaviour of the bias against dimension was the same for the diagonal and factor structure cases. 


\subsection{Exploratory Data Analysis}\label{appssec:eda}
The cross-sectional average of mean returns and standard deviations is $2.7\%$ p.a and $33\%$ p.a, respectively. The average lag-1 autocorrelation across assets is approximately $0.08$. Notably, 4 of the 153 assets display statistically significant negative autocorrelation. The returns display some factor structure, with the first five principal components explaining $80\%$ of the variation in returns, and the first thirteen principal components explaining $90\%$ of the variation. Although these are factor returns, the ``meta-factor" structure likely arises since many factors are proxies for the same underlying risk factor. We note that the parameter ranges change over different sub-periods, typically displaying more cross-sectional variation than over the whole period. The exploratory plots of the unobservable parameters show that the autocorrelation in the mean process $\phi$ has been specified to have a linear relationship with the autocorrelation in the returns $\psi$, where the dispersion in $\phi$ for a given level of $\psi$ arises from differences in the ratio $\tfrac{\sigma_
\mu}{\sigma_R^2}$. The cross-sectional mean SVR is $0.39$. 

\end{document}